\documentclass{IEEEojcsys}

\usepackage[colorlinks,urlcolor=blue,linkcolor=blue,citecolor=blue]{hyperref}
\usepackage{color,array}
\usepackage{graphicx}
\usepackage{mathrsfs}
\usepackage{pifont}
\newif\ifdouble
\doublefalse

\usepackage[numbers]{natbib}
\usepackage{xurl}

\usepackage{amsmath}
\usepackage{amsthm}
\usepackage{amssymb}
\usepackage{resizegather}
\usepackage{xcolor}
\usepackage{soul}
\usepackage{subcaption}
\usepackage{todonotes}

\newtheorem{definition}{Definition}

\newtheorem{lemma}{Lemma}
\newtheorem{theorem}{Theorem}
\newtheorem{example}{Example}

\usepackage{booktabs} 

\usepackage{tikz}
\usetikzlibrary{automata,positioning}
\usetikzlibrary{shapes,arrows}
\tikzset{state/.style={circle, draw, minimum size=0.5cm, initial distance=0.2cm, initial text=},}

\usepackage{algorithm}
\usepackage{algorithmicx}
\usepackage{algpseudocode}
\algnewcommand{\LineComment}[1]{\Statex \(\triangleright\) #1}
\makeatletter
\newcommand{\algmargin}{\the\ALG@thistlm}
\makeatother
\newlength{\whilewidth}
\algdef{SE}[parWHILE]{parWhile}{EndparWhile}[1]
  {\parbox[t]{\dimexpr\linewidth-\algmargin}{%
     \hangindent\whilewidth\strut\algorithmicwhile\ #1\ \algorithmicdo\strut}}{\algorithmicend\ \algorithmicwhile}%
\algnewcommand{\parState}[1]{\State%
  \parbox[t]{\dimexpr\linewidth-\algmargin}{\strut #1\strut}}
\algnewcommand{\parRequire}[1]{\Require%
  \parbox[t]{\dimexpr\linewidth-\algmargin}{\strut #1\strut}}

\jvol{00}
\jnum{XX}
\paper{1234567}
\pubyear{2021}
\receiveddate{XX September 2021}
\accepteddate{XX October 2021}
\publisheddate{XX November 2021}
\currentdate{XX November 2021}
\doiinfo{OJCSYS.2021.Doi Number}

\newcommand{\nat}{\mathbb{N}}

\renewcommand{\subset}{\subseteq }

\renewcommand{\epsilon}{\varepsilon}

\renewcommand{\epsilon}{\varepsilon}
\renewcommand{\c}{\circ}

\newcommand{\rr}{\mathcal{R}}

\newcommand{\attacker}{\mathcal{A}}
\newcommand{\plant}{\mathcal{P}}
\newcommand{\supervisor}{\mathcal{S}}
\newcommand{\desired}{\mathscr{K}}

\newcommand{\astate}{\mathsf{s}}
\newcommand{\astates}{\mathsf{S}}
\newcommand{\ainit}{\mathsf{s}_\mathrm{init}}
\newcommand{\atransitions}{\mathsf{Trans}}

\newcommand{\ainput}{\mathsf{i}}
\newcommand{\ainputs}{\mathbf{I}}
\newcommand{\aoutput}{\mathsf{o}}
\newcommand{\aoutputs}{\mathbf{O}}

\newcommand{\afinal}{\mathsf{S}_\mathrm{final}}

\newcommand{\abs}[1]{\vert #1 \vert}

\newcommand\old[1]{{\color{gray} #1}}

\renewcommand\old[1]{}

\newcommand*{\formatsection}{\Roman{section}}
\newcommand*{\formatsubsection}{\Alph{subsection}}
\newcommand*{\formatsubsubsection}{\arabic{subsubsection}}
\renewcommand*{\thesection}{\formatsection}
\renewcommand*{\thesubsection}{\thesection-\formatsubsection}
\renewcommand*{\thesubsubsection}{\thesubsection.\formatsubsubsection}
\begin{document}

\sptitle{Formal Verification and Synthesis of Cyber-Physical Systems}

\title{Attack-Resilient Supervisory Control of Discrete-Event Systems: A Finite-State Transducer Approach} 

\editor{This paper was recommended by Associate Editor XXX.}

\author{Yu Wang\affilmark{1}}
\author{Alper Kamil Bozkurt\affilmark{2}}
\author{Nathan Smith\affilmark{1}}
\author{Miroslav Pajic\affilmark{2}}
\affil{Department of Mechanical \& Aerospace Engineering, University of Florida, Gainesville, FL 32611, USA.} 
\affil{Department of Electrical \& Computer Engineering, Duke University, Durham, NC 27710, USA.} 

\corresp{CORRESPONDING AUTHOR: Yu Wang (e-mail: \href{mailto:yuwang1@ufl.edu}{yuwang1@ufl.edu})}

\authornote{This work was partly supported by the awards ONR N00014-17-1-2012 and N00014-17-1-2504, AFOSR FA9550-19-1-0169, and NSF CNS-1652544.}

\markboth{PREPARATION OF PAPERS FOR IEEE OPEN JOURNAL OF CONTROL SYSTEMS}{WANG {\itshape ET AL}.}

\begin{abstract}
Resilience to sensor and actuator attacks is a major concern in the supervisory control of discrete events in cyber-physical systems (CPS). In this work, we propose a new framework to design supervisors for CPS under attacks using finite-state transducers (FSTs) to model the effects of the discrete events. FSTs can capture a general class of regular-rewriting attacks in which an attacker can nondeterministically rewrite sensing/actuation events according to a given regular relation. These include common insertion, deletion, event-wise replacement, and finite-memory replay attacks. We propose new theorems and algorithms with polynomial complexity to design resilient supervisors against these attacks. We also develop an open-source tool in Python based on the results and illustrate its applicability through a case study.
\end{abstract}

\begin{IEEEkeywords}
cyber-physical systems, formal methods, control system security
\end{IEEEkeywords}

\maketitle

\section{Introduction}
\label{sec:intro}

Supervisory control is a widely-used high-level control technique to deal with discrete events (e.g., turning on/off one of many switches) in cyber-physical systems (CPS) that work for various applications including transportation~\cite{shepard12,pajic_csm17}, smart infrastructure~\cite{cassandras_SmartCitiesCyberPhysical_2016}, and healthcare~\cite{jiang_pieee12}. The goal of the supervisor, typically implemented by cyber controllers, is to ensure that the possible discrete events happen in the correct sequences to prevent system failure. Mathematically, the discrete events are captured by a set of symbols that are implemented by sensors and actuators, their effect on the controlled physical plant is captured by finite-state machines whose transitions are driven by these symbols, and the goal is to apply feedback control to restrict the plant's execution, as shown in Fig.~\ref{fig:diagram}.

With increasing applications in contested scenarios, such as autonomous driving~\cite{raman_ReactiveSynthesisSignal_2015,pajic_csm17} and distributed manufacturing~\cite{jakovljevic2019distributing,lesi2020security}, there is a growing interest in ensuring the resilience of supervisors against sensor/actuator attacks~\cite{lin2019towards,lima2019security,meira2021synthesis}. In supervisory control, the possible attacks are illustrated in Fig.~\ref{fig:diagram}. The sensor attacks aim to corrupt the true sensing symbols from the plant to the supervisor by either hacking into the plant's sensor~\cite{spoofedgps,warner2002simple,shoukry2013non} or the communication network~\cite{smith_decoupled_attack,ncs_attack_models,jovanov_tac19,lesi_tecs17}. Similarly, the actuator attacks aim to corrupt the true actuating symbols from the supervisor to the plant by either hacking into the plant's actuator or the communication network. This setup captures simultaneous (including coordinated) attacks on sensors and actuators and is more general than~\cite{wakaiki_SupervisoryControlDiscreteevent_2017,carvalho_DetectionMitigationClasses_2018} for either sensor or actuator attacks.

\begin{figure}[!t]
\usetikzlibrary{arrows}
\centering
\begin{tikzpicture}
\draw[black] (0,0) rectangle (6,.5);
\draw [dotted] (1.5, 0) -- (1.5, 0.5);
\draw [dotted] (4.5, 0) -- (4.5, 0.5);
\draw[black] (1.5,2) rectangle (4.5,2.5);
\draw [dotted, rounded corners] (0, 1) rectangle (6,1.5);

\draw [->, >=latex] (0.75, 0.5) -- (0.75,1.35);
\draw (0.75,1.3) -- (0.75,2.25);
\draw (0.75,2.25) -- node [above] {$\ainput_{\supervisor}$} (1.5,2.25);
\draw (4.5,2.3) -- node [above] {$\aoutput_{\supervisor}$} (5.25,2.25);
\draw [->, >=latex] (5.25,2.25) -- (5.25,1.2);
\draw (5.25,1.25) -- (5.25,0.5);

\draw [->,dashed, >=stealth] (6,2.5) -- (5.35, 1.5);
\draw [->,dashed, >=stealth] (6,2.5) --  (5.35, 0.5);

\draw [->,dashed, >=stealth] (0,2.5) --  (0.65, 1.5);
\draw [->,dashed,  >=stealth] (0,2.5) -- (0.65, 0.5);

\node at (3,0.25) {Plant};
\node at (5.25,0.25) {Actuator};
\node at (0.75,0.25) {Sensor};
\node at (3,1.25) {Communication Network};
\node at (3,2.25) {Supervisor};

\node[align=center] at (6,3) {\small Actuator \\ \small Attacks};
\node[align=center] at (0,3) {\small Sensor \\ \small Attacks};
\end{tikzpicture}
\caption{
Supervisory control of cyber-physical systems. At each time, the plant sends a sensing symbol, which will be corrupted by  $\aoutput_{\plant}$ to the supervisor. Then, the supervisor replies to the plant with an actuating symbol $\ainput_{\plant}$. A plant transition can only happen if the pair $(\ainput_{\plant},\aoutput_{\plant})$ matches the transition's label. In practice, the symbols $\aoutput_{\plant}$ and $\ainput_{\plant}$ may be revised by malicious attackers, causing the plant to make transitions not allowed by the supervisor.} \label{fig:diagram}  
\end{figure}

The block diagram for supervisory control under sensor/actuator attacks is shown in Fig.~\ref{fig:control}. In a real system, multiple sensor attacks can happen via different vectors (e.g., network or sensor). In Fig.~\ref{fig:control}, the overall effect is represented by a single attacker $\attacker_s$. The input-output relation of $\attacker_s$ captures how a sequence of true sensing symbols may be nondeterministically rewritten into a sequence of corrupted sensing symbols. Similarly, the overall effect of multiple possible actuator attacks is represented by a single attacker $\attacker_a$, whose input-output relation captures how a sequence of true actuator symbols may be nondeterministically rewritten into a sequence of corrupted actuator symbols.

The common presence of attackers in CPS has motivated recent works on developing new theories and algorithms for attack-resilient supervisory control. However, 
many of them only deal with simple and history-independent attack strategies, e.g., symbol insertion, deletion, or replacement attacks (e.g.,~\cite{wakaiki_SupervisoryControlDiscreteevent_2017}), while many attackers in CPS can use complex and history-dependent attack strategies. Among the works involving history dependent attacks (e.g., ~\cite{meiragoes2020synthesis}), the attacks are typically modeled by a function: the attack is determined by the plant's transition history. Our attacker model is described by an FST giving an intuitive visual representations of the attackers.

Take the replay attack~\cite{stuxnet,stuxnet2} as an example, which is a common strategy for launching cyber-attacks. It works by first recording a fragment of symbols and then replaying it repeatedly to trap the system. Implementing this attack requires the attacker to decide when to continue recording or start replaying. This requires the attackers to possess internal states to make such decisions based on previous actions. The idea of using states to model attack behaviors has been proposed in~\cite{meiragoes2020synthesis}. However, in that work, the state space is shared between the attack and the system. In practice, the attacker and the system are typically implemented separately. Thus, it is more appropriate to model the attacker individually with its own state space. This will help study the impact of different attacks on the same plant and study the impact of multiple combined attacks.


This work proposes to use finite-state transducers (FSTs)~\cite{sakarovitch_ElementsAutomataTheory_2003,droste_HandbookWeightedAutomata_2009}, which generalize finite-state automata, to model the complex and history-dependent strategies of the attackers. 
These FST models can be viewed as abstractions of cyber/network attacks implemented by embedded programs (e.g., malware).
FST transitions are driven by an input symbol and also produce a different output. This feature allows FSTs to capture general complex attack strategies, including previously-studied attacks (e.g., symbol insertion, deletion, or replacement) and new history-dependent attacks (e.g., replay attacks). In addition, one can easily compose multiple attack strategies to form new attack strategies via FST model compositions~\cite{mohri_FiniteStateTransducersLanguage_1997,mohri_WeightedFinitestateTransducers_2002,mohri_WeightedFiniteStateTransducer_2004,mohri_WeightedAutomataAlgorithms_2009}, which have polynomial complexity and are well supported by existing C\texttt{++}/Python libraries (e.g.,~\cite{allauzen_OpenFstGeneralEfficient_2007}). Finally, one can implement constraints on unconstrained attack strategies and facilitate problem analysis with common security measures (e.g., intrusion detectors~\cite{bozkurt2020secure}, intermittently authentication~\cite{jovanov_tac19,lesi_rtss17,wang2019attack}).

Our main contribution is the development of a new theory to synthesize resilient supervisors against sensor and actuator attacks. The supervisory control diagram is shown in Fig.~\ref{fig:control}. We assume the FST models for attackers are known a priori and capture all the possible attack behaviors (e.g., nondeterministic symbol deletion or insertion). The supervisor is resilient in the sense that it can restrict the plant's execution to an allowed set under the attacks. Our theory gives a constructive algorithm with polynomial complexity to synthesize the supervisor, if feasible. It also shows that the feasible supervisor is realizable by an FST. 


This work improves our previous work~\cite{wang_AttackResilientSupervisoryControl_2019} in two aspects. First, we generalize it to plants modeled by FSTs (instead of automata) to handle CPS equipped with sensors and actuators that yield different input and output symbols. Our plant model is similar to~\cite{xu2009discrete,ushio2015nonblocking}, although no attacks were considered in those works. Second, we develop an open-source tool in Python based on the new results and illustrate its applicability through a case study.
The rest of the paper is organized as follows. We provide preliminaries on FSTs and regular relations in Section~\ref{sec:prelim} and formalize the problem in Section~\ref{sec:formulation}. Then, we demonstrate the advantages of using FSTs to model attacks in Section~\ref{sec:attack} and provide resiliency conditions and algorithms to design resilient supervisors for FST-based sensor and actuator attacks in Section~\ref{sec:attack}. Finally, case studies are presented in Section~\ref{sec:sim}, before concluding in Section~\ref{sec:conclusion}.

\paragraph*{Related work}
Supervisory control under attacks has been studied extensively in the literature, so we provide a detailed comparison with previous work as follows. This work considers simultaneous actuator and sensor attacks, while only sensor attacks are considered in~\cite{su2018supervisor}. Furthermore, our attack model allows symbol revision of unbounded length, while~\cite{su2018supervisor} only studied bounded attacks. We notice that~\cite{lima2022security,you2022livenessenforcing} consider simultaneous actuator and sensor attacks. Our work can handle regular desired languages while~\cite{lima2022security} only focuses on reachability. In addition, we provide an explicit attack model via FST, while the attack model is only given implicitly in~\cite{lima2022security} through the attacked plant model. Our attack model can be history-dependent and thus more complex than the history-independent attack model in~\cite{you2022livenessenforcing}. In addition, we deal with the regular desired languages instead of liveness in~\cite{you2022livenessenforcing}. The problem of synthesizing complex attackers has been studied in ~\cite{goes_StealthyDeceptionAttacks_2017, lin2020synthesis, lin2021synthesis, meiragoes2020synthesis, tai2023synthesis, yao2022sensor}, while our work focuses on the supervisor synthesis problem. Our work can synthesize supervisors to counter the attacks, while~\cite{zhang2021joint} only focuses on synthesizing estimators to detect the attacks. Our work can handle attackers that tamper with the system dynamics, while~\cite{tai2023privacypreserving} can only handle eavesdroppers. Compared to~\cite{meira2021synthesis} that deals with supervisor design under attacks, we focus on restricting the plant to a desired language $\desired$, while the supervisor in \cite{meira2021synthesis} prevents the plant from reaching unsafe states. Consequently, their supervisor synthesis problem is converted to an equivalent game theoretic dual problem. Lastly, our framework handles both sensor and actuator attackers simultaneously while \cite{meira2021synthesis} is restricted to only sensor attackers.

\section{Preliminaries on FSTs}
\label{sec:prelim}

To start with, we introduce the following common notations for arithmetic over symbols. Specifically, we denote the set of natural numbers including zero by $\nat$, set cardinality by $\abs{S}$, power set by $2^S$, and set subtraction by $S_1 \backslash S_2 = \{s \in S_1 \mid s \notin S_2\}$. We write ``iff'' for ``if and only if''. For $n \in \nat$, let $[n] = \{1,..,n\}$. We follow the common notations for sequences. For a given finite set of symbols, a finite-length sequence of symbols is called a word. A set of words is called a language. A word can be viewed as a singleton language. The empty symbol/word is denoted by $\epsilon$. The concatenation of two languages (or words) is defined by
$$L_1 L_2 = \{I_1 I_2 \ \vert \ I_1 \in L_1, I_2 \in L_2\}.$$
The Kleene star (i.e., finite repetition) of a language (or word) is defined by 
$L^* = \{I_1 \ldots I_n \mid I_1, \ldots, I_n \in L, n \in \nat\}$; by convention, $\varepsilon \in L^*$. 
For singleton sets, the notation convention is $I_1^* =\{I_1\}^*$. The union of two languages (or words) is the same as the union of sets, denoted by $L_1 \cup L_2$. 
A language is \emph{regular} if it can be represented only using union, concatenation, and Kleene star~\cite{sipser_IntroductionTheoryComputation_1996}.
In addition, a word $I_1$ is a prefix of another word $I$ if there exists $I_2$ such that $I = I_1 I_2$. The prefix closure $\overline{L}$ of a language $L$ is derived by including all prefixes of all $I \in L$ in $\overline{L}$. A language $L$ is prefix-closed if $L =  \overline{L}$.

Now, we provide a mathematical introduction to FSTs. FSTs can be seen as automata with input and output symbols on their transitions. Similarly, automata can be viewed as FSTs with identical input and output.

\begin{definition}
\label{def:fst}
A finite-state transducer (FST) is a tuple $\attacker = (\astates, \ainit, \ainputs, \aoutputs, \atransitions, \allowbreak \afinal)$ where
\begin{itemize}
  \item $\astates$ is a finite set of states;
  \item $\ainit \in \astates$ is the initial state;
  \item $\ainputs$ is a finite set of non-empty input symbols;
  \item $\aoutputs$ is a finite set of non-empty output symbols;
  \item $\atransitions \subseteq \astates \times \big( \ainputs \cup \epsilon \big) \times \big( \aoutputs \cup \epsilon \big) \times \astates$ is a transition relation, where $\varepsilon$ is the  empty symbol;
  \item $(s, \epsilon, \epsilon, s) \in \atransitions$ for all $s \in S$;
  \item $\afinal \subseteq \astates$ is a finite set of final states.
\end{itemize}
\end{definition}  

In Definition~\ref{def:fst}, the symbol $\epsilon$ stands for the empty symbol. An FST transition with $\epsilon$ as the input symbol can self-trigger. An FST transition with $\epsilon$ as the output symbol yields no output symbol. The self-loop $(s, \epsilon, \epsilon, s)$ means the FST can stay in the same state without receiving an input symbol and generating an output symbol. 

\paragraph*{Languages of an FST}
For the FST $\attacker$, we define an \emph{execution} by a sequence of transitions $(\astate_0, \ainput_1, \aoutput_1, \astate_1) \allowbreak ... (\astate_{n-1}, \ainput_n, \aoutput_n, \astate_n)$ where $\astate_0 = \ainit$ and $(\astate_{i-1}, \ainput_i, \aoutput_i, \astate_i) \allowbreak \in \atransitions$ for $i \in [n]$. The execution defines an allowed word of input/output pairs $(\ainput_1, \aoutput_1) \ldots (\ainput_n, \aoutput_n)$. The set of such allowed words (i.e., sequences of input/output symbol pairs) is called the \emph{language} of the FST $\attacker$, denoted by $L(\attacker)$. In addition, we call $I = \ainput_1 \ldots \ainput_n$ an allowed input word and $O = \aoutput_1 \ldots \aoutput_n$ an allowed output word. The set of allowed input words is the \emph{input language} of $\attacker$, denoted by $L_{\mathrm{in}}(\attacker)$. Similarly, the \emph{output language} is defined and denoted by $L_{\mathrm{out}}(\attacker)$.

\paragraph*{FSTs and Relations}
The FST $\attacker$ defines a \emph{relation} between $\ainputs^*$ and $\aoutputs^*$ by 
\begin{align*}
& \rr_\attacker = \{(\ainput_1 \ldots \ainput_n, \aoutput_1 \ldots \aoutput_n) \ \vert \ \exists \textrm { an execution }  (\ainit, \ainput_1, \aoutput_1, \astate_1) \\ & ... (\astate_{n-1}, \ainput_n, \aoutput_n, \allowbreak \astate_n) \textrm{ and } \astate_n \in \afinal\} \subseteq \ainputs^* \times \aoutputs^*
\end{align*}
In this case, we say the FST $\attacker$ realizes the relation $\rr_\attacker$. More specifically, $\rr_\attacker$ can be seen as a relation between the input language $L_{\mathrm{in}}(\attacker)$ and the output language $L_{\mathrm{out}}(\attacker)$, since $L_{\mathrm{in}}(\attacker) = \rr_\attacker^{-1} (\aoutputs^*)$ and $L_{\mathrm{out}}(\attacker) = \rr_\attacker (\ainputs^*)$. Here, $\rr_\attacker$ is not necessarily a function since an input word can be mapped nondeterministically to multiple output words.


To facilitate further discussion on relations, we introduce the following common notations. 
For a relation $\rr \subseteq  S_1 \times S_2$, we define (with a slight abuse of notation) the image of a subset $T_1 \subseteq  S_1$ (an element is viewed as a singleton set) over the relation $\rr$ by 
$$\rr(T_1) = \{s_2 \in S_2 \mid s_1 \in T_1, (s_1, s_2) \in \rr \}.$$ The relation $\rr$ is a partial function if $\abs{\rr(s_1)} \leq 1$ for any $s_1 \in S_1$. The inverse of the relation $\rr$ is defined by 
$$\rr^{-1} = \{(s_2, s_1) \mid (s_1, s_2) \in \rr \},$$ 
whereas the composition of two relations is defined by 
$$\rr_1 \circ \rr_2 = \{(s_1, s_3) \mid \exists s_2. \ (s_1, s_2) \in \rr_1, (s_2, s_3) \in \rr_2\}.$$ Here, the relation composition is read from left to right.

\paragraph*{FSTs and Automata}
Like finite automata define regular languages, FSTs define \emph{regular relations} as described below.

\begin{definition}
\label{def:regular relation}
For two finite sets $\ainputs$ and $\aoutputs$, the relation $\rr \subseteq \ainputs^* \times \aoutputs^*$ is a regular relation iff 
$$\big\{ (\ainput_1, \aoutput_1) \ldots (\ainput_n, \aoutput_n) \mid  (\ainput_1 \ldots \ainput_n, \aoutput_1 \ldots \aoutput_n) \in \rr \big\}$$
is a regular language over $\ainputs \times \aoutputs$. Here, $\ainput_1, \aoutput_1, \ldots \ainput_n,  \aoutput_n$ can be the empty symbol $\varepsilon$.
\end{definition}

Since FSTs can be seen as automata with input and output symbols on their transitions, we have the following lemma~\cite{holcombe_AlgebraicAutomataTheory_1982}.

\begin{lemma} 
\label{lem:1}
The relation defined by an FST is regular. Any regular relation is realizable by an FST.
\end{lemma}




\section{Problem Formulation} 
\label{sec:formulation}

This section introduces our problem formulation for attack-resilient supervisory control. Section~\ref{sub:formulation-1} discusses how to properly use FST models in supervisory control. Section~\ref{sub:formulation-2} presents the fundamentals of the supervisory control of plants modeled by FSTs without attacks. Section~\ref{sub:formulation-3} discusses the supervisory control with attackers modeled by~FSTs.

\subsection{Using FST models in supervisory control}
\label{sub:formulation-1}

We use the FSTs from Definition~\ref{def:fst} to model the plant in Figure~\ref{fig:control}. Since the FSTs are placed in a control loop, the transitions are implicitly timed. The FSTs are executed iteratively, and each iteration requires a unit of time. 

\begin{definition} 
\label{def:observable}
An FST $\attacker$ is observable if 1) it has no transition labeled by $(\varepsilon, \varepsilon)$ other than self-loops and 2) only one execution exists for an allowed word in $L(\attacker)$.
\end{definition}

The observability from Definition~\ref{def:observable} ensures that the executions of the FSTs, and thus their current states, are uniquely determined by both the previous input/output words. It can facilitate the hardware and software implementation of the plant and attackers. In addition, the observable FSTs from Definition~\ref{def:observable} allow empty input or output symbols. In the supervisory control setup, an empty symbol means the plant or attackers give no input or output for the current time. 

The observability described in Definition~\ref{def:observable} strict the transition in the FST Definition~\ref{def:fst} from a relation to a (partial) function $\atransitions:\astates \times \ainputs \times \aoutputs \rightarrow \astates$. An observable FST is different from a Mealy automaton whose transition is defined as the (partial) function $\atransitions:\astates \times \ainputs \rightarrow \astates \times \aoutputs$~\cite{mealy1955method}. Thus, observable FSTs can still capture nondeterministic attack behaviors that rewrite an input symbol into different output symbols. We will discuss these issues in detail in Section~\ref{sec:attack}.

Two words $(\ainput_1, \aoutput_1) \ldots (\ainput_n, \aoutput_n)$ and $(\ainput_1', \aoutput_1') \ldots (\ainput_m', \aoutput_m')$ of the plant or attackers are viewed as different even if $\ainput_1 \ldots \ainput_n = \ainput_1' \ldots \ainput_m'$ and $\aoutput_1 \ldots \aoutput_n = \aoutput_1' \ldots \aoutput_m'$ after ignoring the empty symbols. More specifically, for a plant, $(\ainput_1, \varepsilon) (\varepsilon, \aoutput_1)$ and $(\ainput_1, \aoutput_1) (\varepsilon, \varepsilon)$ are different words. The former means the plant inputs $\ainput_1$ and outputs nothing at Time $1$ and inputs nothing and outputs $\ainput_2$ at Time $2$; the latter means the plant inputs $\ainput_1$ and outputs $\ainput_2$ at Time $1$ and inputs and outputs nothing at Time $2$.

An FST transition labeled with input/output symbols $(\varepsilon, \varepsilon)$ is similar to the $\varepsilon$-transitions in automata. They can happen spontaneously without receiving and generating any input and output. In addition, observing the FST's input and output cannot determine whether they have happened or not. Such unobservability is unrealistic and undesirable in modeling plants and attackers in the control loop. Thus, we focus on using observable FSTs in this work.  

\subsection{Supervisory control of FST plants}
\label{sub:formulation-2}


\begin{figure}[h]
\centering
\begin{tikzpicture}
\node (center) {};

\node [draw, below of=center, node distance=0.7cm]  (p) {Plant $\plant$};
\node [draw, above of=center, node distance=0.7cm] (s) {Supervisor $\mathcal{S}$};

\draw [->] (s.east) -- ++(0.3,0) |- node [right] {$\ainput_{\plant} = \aoutput_{\supervisor}$} (p.east);
\draw [->] (p.west) -- ++(-0.7,0) |- node [left] {$\aoutput_{\plant} = \ainput_{\supervisor} $} (s.west);
\end{tikzpicture}

\caption{Block diagram for supervisory control of plants modeled by FSTs. 
} \label{fig:control}
\end{figure}

The plant modeled by FSTs can generate output symbols that are different from the input symbols. Thus, unlike classic supervisory control \cite{cassandras_IntroductionDiscreteEvent_2008} using automata as supervisors, we use FSTs as supervisors to control such plants, as shown in Figure~\ref{fig:control}. Accordingly, the control loop executes iteratively as follows. Both the supervisor and the plant start from their initial states. 
    
\begin{enumerate}
    \item The supervisor (modeled by an FST) chooses an output symbol $\aoutput_{\supervisor}$, which will be sent to the plant, based on an eligible transition that has this symbol as its output.  
    \item The plant receives $\ainput_{\plant} = \aoutput_{\supervisor}$ and makes a transition non-deterministically whose input symbol matches it. This transition will generate an output symbol $\aoutput_{\plant}$.
    \item Upon receiving $\ainput_{\supervisor} = \aoutput_{\plant}$, the supervisor completes this iteration by executing the transition labeled by $(\ainput_{\supervisor}, \aoutput_{\supervisor}) = (\aoutput_{\plant}, \ainput_{\plant})$. Otherwise, when there is no such transition, the supervisor will stop the control loop and raise an alarm.  
\end{enumerate}

In this setup, the supervisor has the freedom to choose the symbol $\aoutput_{\supervisor}$ in each iteration. Upon receiving $\ainput_{\plant} = \aoutput_{\supervisor}$, the plant has the freedom to choose one of many transitions whose input symbol matches it.
The set of all possible plant executions in the supervisory control loop can be captured by an FST derived from the loop composition defined below. 

\begin{definition}
\label{def:loop_composition}
The loop composition of the plant $\plant = (\astates_{\plant}, \mathsf{s}_{i,\plant}, \ainputs_{\plant}, \aoutputs_{\plant}, \atransitions_{\plant}, \mathbf{S}_{f,\plant})$ with the supervisor $\supervisor  \allowbreak = (\astates_{\supervisor}, \mathsf{s}_{i,\supervisor}, \ainputs_{\supervisor}, \aoutputs_{\supervisor}, \atransitions_{\supervisor}, \mathbf{S}_{f,\supervisor})$ in Figure~\ref{fig:control} is an FST given by $\plant \vert \supervisor = (\astates_{\plant} \times \astates_{\supervisor}, ( \mathsf{s}_{i,\plant}, \mathsf{s}_{i,\supervisor} ), \ainputs_{\plant}, \aoutputs_{\plant}, \atransitions_{loop}, \allowbreak \mathsf{S}_{f,\plant} \times \mathsf{S}_{f,\supervisor})$  
where the transition $( (\astate_{1,\plant}, \astate_{1,\supervisor}), \ainput_{\plant}, \aoutput_{\plant}, (\astate_{2,\plant}, \astate_{2,\supervisor}) ) \in \atransitions_{loop}$ if the following two conditions are met; 
\begin{enumerate}
    \item $( \astate_{1,\plant}, \ainput_{\plant}, \aoutput_{\plant}, \astate_{2,\plant} ) \in \atransitions_{\plant}$ 
    \item $( \astate_{1,\supervisor}, \aoutput_{\plant}, \ainput_{\plant}, \astate_{2,\supervisor} ) \in \atransitions_{\supervisor}$. 
\end{enumerate}
\end{definition}

Accordingly, the plant's executions in the supervisory control loop in Figure~\ref{fig:control} is given by $L(\plant \vert \supervisor)$, i.e., the language of the loop composition $\plant \vert \supervisor$. 


\begin{definition}
\label{def:controllability-noattack}
Let $\plant$ be an observable FST plant and $\desired \subseteq L(\plant)$ be the prefix-closed regular language capturing its desired executions. The supervisor $\supervisor$ controls the plant $\plant$ to the language $\desired$ if and only if the set of plant executions under the supervisor's regulation satisfies
\[
L(\plant \vert \supervisor) = \desired.
\]
\end{definition}

\begin{example}
Figure~\ref{fig:ex1} shows an example of the supervisory control of FST plants. Suppose the plant is modeled by the FST shown by Figure~\ref{fig:ex1-plant}. It has one state $0$ and two transitions $t_1$ and $t_2$ with input/output symbols $(\alpha_1, \alpha_2)$ and $(\alpha_2, \alpha_2)$, respectively. The goal is to restrict the plant execution to $\overline{(t_1 t_2)^*}$, or equivalently restrict the plant language to $\desired = \overline{\big( (\alpha_1, \alpha_2) (\alpha_2, \alpha_2) \big)^*}$.
  For this goal, consider a supervisor realized by an FST shown by Figure~\ref{fig:ex1-supervisor}, which has two states $0$ and $1$ and two transitions $\tau_1$ and $\tau_2$. At Time $1$, only the transition $\tau_1$ labeled by $(\alpha_2, \alpha_1)$ is allowed by the supervisor whose current state is $0$. Thus, only the input/input pair $(\alpha_1, \alpha_2)$ is allowed on the plant, and only the transition $t_1$ can happen. The symbols are flipped since the plant's input is the supervisor's output and the plant's output is the supervisor's input. At Time $2$, the supervisor state jumps to $1$ following the transition $\tau_1$. Now, only the transition $\tau_2$ labeled by $(\alpha_2, \alpha_2)$ is allowed by the supervisor. Thus, only the input/output pair $(\alpha_2, \alpha_2)$ is allowed on the plant, and only the transition $t_2$ can happen. If the plant tries to execute the other transition $t_1$, the supervisor will stop the control loop and raise the alarm. Repeating the process, the supervisor restricts the plant's execution to $\overline{(t_1 t_2)^*}$ before stopping. 
\end{example}

\begin{figure}[h]
\centering
\subcaptionbox{Plant $\plant$.\label{fig:ex1-plant}}[.45\linewidth]{
\begin{tikzpicture}
  \node[state, initial, initial text=] (s) {$0$};
  \path[->] (s) edge[loop right] node {$t_1: (\alpha_1, \alpha_2)$} ();
  \path[->] (s) edge[loop above] node {$t_2: (\alpha_2, \alpha_2)$} ();
\end{tikzpicture}
}
\subcaptionbox{Supervisor $\supervisor$.\label{fig:ex1-supervisor}}[.45\linewidth]{
\begin{tikzpicture}
  \node[state,initial, initial text=] (0) {$0$};
  \node[state,right of=0, xshift=0.5cm] (1) {$1$};
  \path[->] (0) edge[bend left] node[above] {$\tau_1: (\alpha_2, \alpha_1)$} (1)
  (1) edge[bend left] node[below] {$\tau_2: (\alpha_2, \alpha_2)$} (0);
\end{tikzpicture}
}
\caption{Example of supervisory control of FST plants.\label{fig:ex1}}
\end{figure}



Based on the loop composition, we introduce the following theorem on the supervisory control of FST plants without attacks.
\begin{theorem} \label{thm:1}
The observable FST $\supervisor$ that has the language
$$L(\supervisor) = 
\big\{ (\aoutput_1, \ainput_1) \ldots (\aoutput_n, \ainput_n) \mid  (\ainput_1, \aoutput_1) \ldots (\ainput_n, \aoutput_n) \in \desired \big\}$$
controls the $\plant$ to the desired language $\desired$.
\end{theorem}

\begin{proof}
The FSTs $\supervisor$ and $\plant$ can be viewed as automata that take input/output symbol pairs. Thus, the theorem follows the classic supervisor control theory for automata \cite{cassandras_IntroductionDiscreteEvent_2008}. Specifically, the FST realization $\supervisor$ of the supervisor exists since $\desired$ is regular. Because $\desired$ is prefix-closed, any state of $\supervisor$ is a final state, i.e., the executions of $\supervisor$ can stop at any time.
Finally, the symbols are flipped since the plant's input is the supervisor's output and the plant's output is the supervisor's input. 
\end{proof}


\subsection{Including attacks in supervisory control}
\label{sub:formulation-3}


Now we consider simultaneous sensor and actuator attacks between the supervisor $\supervisor$ and the plant $\plant$ as shown in Figure~\ref{fig:control_attack}. 

\begin{figure}[!t]
\centering
\begin{tikzpicture}
\node (center) {};
\node [draw, right of=center, node distance=1.5cm]  (ai) {Attacker $\attacker_a$};
\node [draw, below of=center, node distance=1cm]  (p) {Plant $\plant$};
\node [draw, left of=center, node distance=1.5cm] (ao) {Attacker $\attacker_s$};
\node [draw, above of=center, node distance=1cm] (s) {Supervisor $\mathcal{S}$};

\draw [->] (ai) |- node [below,right] {$\ainput_{\plant}$} (p);
\draw [->] (p) -| node [below,left] {$\aoutput_{\plant}$} (ao);
\draw [->] (ao) |- node [above,left] {$\ainput_{\supervisor}$} (s);
\draw [->] (s) -| node [above,right] {$\aoutput_{\supervisor}$}  (ai);

\draw [dashed] (-2.75,-0.5)--(-2.75,1.5) ;
\draw [dashed] (-2.75,1.5)--(2.75,1.5) node[midway, above]{Corrupted Supervisor $\supervisor_c$};
\draw [dashed] (2.75,1.5)--(2.75,-0.5)--(-2.75,-0.5);
\end{tikzpicture}
\caption{Block diagram for supervisory control under sensor and actuator attacks. The serial composition of $\attacker_a$, $\plant$, and $\attacker_s$ in the dashed box form the corrupted supervisor. The executions of the plant $\plant$ in this control loop is derived by its loop composition with the corrupted supervisor.
} \label{fig:control_attack}
\end{figure}

\paragraph*{Attack Model}

In practice, there can be many attacks (e.g., via hacking into sensors or communications) that corrupt the symbols sent between the plant to the supervisor. In this work, we propose to capture their overall effect by two FSTs, the actuator attacker $\attacker_a$ and the sensor attacker $\attacker_s$. The input-output relation of the FSTs $\attacker_a$ and $\attacker_s$ captures how a sequence of true sensing/actuating symbols can be revised nondeterministically into another sequence of corrupted symbols. Mathematically, these relations fall into the category of \emph{regular relations} \cite{holcombe_AlgebraicAutomataTheory_1982}, according to Lemma~\ref{lem:1}. For simplicity, we assume $\attacker_s$ and $\attacker_a$ are observable so their internal executions can be captured by their languages. 

The regular-rewriting attacks provide a general framework to model various attack behaviors. They generalize previous history-independent attacks (e.g., insert, delete, and replace) studied in~\cite{wakaiki_SupervisoryControlDiscreteevent_2017,su_SupervisorSynthesisThwart_2018,goes_StealthyDeceptionAttacks_2017,carvalho_DetectionMitigationClasses_2018} and can capture new history-dependent attacks such as finite-memory replay attacks~\cite{stuxnet,stuxnet2}. Finally, we clarify that attackers modeled by observable FSTs can still rewrite input symbols in different ways to capture nondeterministic attack behaviors, according to Definition~\ref{def:observable}. We will discuss these issues in detail in Section~\ref{sec:attack}.

\paragraph*{Control Loop under Attacks}

With the attackers $\attacker_s$ and $\attacker_a$, the supervisory control loop of Figure~\ref{fig:control_attack}  works as follows. At each time, consider a state transition of the plant with the input and output symbols $\ainput_{\plant}$ and $\aoutput_{\plant}$. Due to the attacks, the plant's true input and output symbols may be revised to corrupted symbols $\ainput_{\plant}$ and $\ainput_{\supervisor}$, respectively, before the output symbol is sent to the supervisor. The supervisor has no access to the true output symbol from the plant and can only see the corrupted symbol. The plant's transition is allowed if and only if the corrupted symbol pair $(\ainput_{\supervisor}, \aoutput_{\supervisor})$ is allowed by the supervisor. Otherwise, the supervisor will stop the control loop and raises the alarm. This procedure is formalized with the following steps:
\begin{enumerate}
    \item The supervisor (modeled by an FST) chooses an output symbol $\aoutput_{\supervisor} $, which will be sent to the plant, based on an eligible transition that has this symbol as its output.  
    \item The actuator attacker $\attacker_a$ receives $\aoutput_{\supervisor}$ and makes a transition whose input symbol matches $\aoutput_{\supervisor}$. There may be many such transitions and the attacker can choose anyone non-deterministically. This transition will generate an output symbol $\ainput_{\plant}$ which will be sent to the plant.
    \item The plant receives $\ainput_{\plant}$ and non-deterministically makes a transition whose input symbol matches it. This transition will generate an output symbol $\aoutput_{\plant}$.
    \item The sensor attacker $\attacker_s$ receives $\aoutput_{\plant}$ and non-deterministically makes a transition whose input symbol matches it. This transition will generate an output symbol $\ainput_{\supervisor}$, which will be sent to the supervisor.
    \item Upon receiving $\ainput_{\supervisor}.$, the supervisor completes this iteration by executing the transition labeled by $(\ainput_{\supervisor}, \aoutput_{\supervisor}) = (\aoutput_{\plant}, \ainput_{\plant})$. Otherwise, when there is no such transition, the supervisor will stop the control loop and raise an alarm.  
\end{enumerate}

Mathematically, the attacker FSTs are joined into the control loop with serial composition. Each attacker is serially composed with the supervisor to create the corrupted supervisor $\supervisor_c$. We now describe how serial composition can be done.

\begin{definition}
\label{def:serial_comp}    
Let $\attacker = (\astates, \allowbreak \ainit, \ainputs, \aoutputs, \allowbreak \atransitions, \afinal)$ and $\attacker' = (\astates', \ainit', \ainputs', \aoutputs', \atransitions', \allowbreak \afinal')$ be two FSTs. The serial composition $\attacker'' = \attacker \circ \attacker'$ is derived by connecting the FSTs $\attacker$ and $\attacker'$ in series as shown in Fig.~\ref{fig:serial}, where the input word sequentially passes $\attacker$ and $\attacker'$. The composition is defined as the FST given by $\attacker'' = (\astates \times \astates', (\ainit, \ainit'), \ainputs, \aoutputs',\atransitions'', \afinal \times \afinal' )$. A transition $( (s_1, s_1'), i, o', (s_2, s_2') )$ if there exists an $\alpha \in \aoutputs \cap \ainputs'$ with the following properties:
\begin{enumerate}
    \item $(s_1, i, \alpha, s_2 )$ is a valid transition in $\atransitions$. 
    \item $( s_1', \alpha, o', s_2')$ is a valid transition in $\atransitions'$. 
\end{enumerate}
\end{definition}

\begin{lemma}
\label{lem:fst-composition}
For any two FSTs $\attacker_1$ and $\attacker_2$, it holds that $\rr_{\attacker_1} \circ \rr_{\attacker_2} = \rr_{\attacker_1 \circ \attacker_2}$ where $\rr$ denotes the regular relation defined by an FST.
\end{lemma}

The model derived from the series composition is still an observable FST. Since FSTs define regular relations, Definition~\ref{def:serial_comp} also provides an FST realization for the composition of regular relations in Lemma~\ref{lem:fst-composition} \cite{holcombe_AlgebraicAutomataTheory_1982}. We can now explicitly define the corrupted supervisor as $\supervisor_c = \attacker_s \circ \plant \circ \attacker_a$. 

\begin{example}
Consider two FSTs $\attacker_1$ and $\attacker_2$ shown in Fig.~\ref{fig:trans_1} and~\ref{fig:trans_2}, where one replaces $\alpha_1$ with $\alpha_2$ and the other injects $\alpha_3$. The overall FST attack model, captured with the serial composition of the two FSTs is derived using Definition~\ref{def:serial_comp} and shown in Fig~\ref{fig:comp}. Also, it holds that $\rr_{\attacker_1 \circ \attacker_2} = \rr_{\attacker_1} \circ \rr_{\attacker_2}$.
\begin{figure}[!t]
\centering
\subcaptionbox{FST $\attacker_1$ \label{fig:trans_1}}[.45\linewidth]{
\begin{tikzpicture}
  \node[state, initial, initial text=] (0) {$0$};
  \node[state, right of=0] (1) {$1$};
  \path[->] (0) edge[bend left] node[above] {$(\alpha_1 , \alpha_2)$} (1);
  \path[->] (1) edge[loop right] node[right] {$(\alpha_1 , \epsilon)$} (1);
\end{tikzpicture}
}
\subcaptionbox{FST $\attacker_2$\label{fig:trans_2}}[.45\linewidth]{
\begin{tikzpicture}
  \node[state, initial, initial text=] (0) {$0$};
  \node[state, right of=0] (1) {$1$};
  \path[->] (0) edge[bend left] node[above] {$(\epsilon , \alpha_3)$} (1);
  \path[->] (0) edge[bend right] node[below] {$(\alpha_2 , \alpha_1)$} (1);
\end{tikzpicture}
}

\par\bigskip

\subcaptionbox{Serial composition $\attacker_1 \circ \attacker_2$\label{fig:comp}}[.5\linewidth]{
\begin{tikzpicture}
  \node[state, initial, initial text=] (0) {$00$};
  \node[state, right of=0, node distance=2cm] (1) {$10$};
  \node[state, below of=0, node distance=1.5cm] (2) {$01$};
  \node[state, right of=2, node distance=2cm] (3) {$11$};

  \path[->] (0) edge node[right] {$(\alpha_1 , \alpha_1)$} (3);
  \path[->] (1) edge[loop right] node[right] {$(\alpha_1 , \epsilon)$} (1);
  \path[->] (3) edge[loop right] node[right] {$(\alpha_1 , \epsilon)$} (3);
  \path[->] (0) edge[bend right] node[left] {$(\epsilon , \alpha_3)$} (2);
  \path[->] (1) edge[bend left] node[right] {$(\epsilon , \alpha_3)$} (3);
\end{tikzpicture}
}
\caption{Example for serial composition of FSTs\label{fig:composition}.}
\end{figure}
\end{example}


\paragraph*{Control Goal}

Let $\desired$ be a desired prefix-closed regular sub-language $\desired \subseteq L(\plant)$ for the plant. The attackers aim to ``stealthily'' incur an undesired plant execution that yields a word outside $\desired$ without triggering the supervisor's alarm. 

We assume that the attackers have some knowledge about the plant. We will now describe why this assumption is made with several simple scenarios where the supervisor can detect an attack, stop the control loop, and raise the alarm. If the actuator attacker sends $\ainput_\plant$ to the plant while the plant currently has no possible transition defined by $\ainput_\plant$ as an input symbol, then the plant will be at a standstill. The supervisor will recognize this abnormality and raise the alarm. This would occur when $L_{\mathrm{out}}(\attacker_a) \not\subset L_{\mathrm{in}}(\plant)$. 

Similarly, if the sensor attacker intercepts a message $\aoutput_{\plant}$ and has no valid transitions with $\aoutput_{\plant}$ as an input symbol in its current state then the supervisor will notice the delay and raise the alarm. In other words, if the supervisor is expecting a response from the plant and it does not receive one then it will raise the alarm and stop the control loop. This failed attack occurs if $L_{\mathrm{out}}(\plant) \not\subset L_{\mathrm{in}}(\attacker_s)$. For the rest of the analysis, we will assume that the attackers have enough knowledge of the plant so that $L_{\mathrm{out}}(\attacker_a) \subset L_{\mathrm{in}}(\plant)$ and $L_{\mathrm{out}}(\plant) \subset L_{\mathrm{in}}(\attacker_s)$.

For the sensor attacker $\attacker_s$, at some time, its input symbol $\aoutput_{\plant}$ may be disallowed by the supervisor in its current state, or its output symbol $\aoutput_{\supervisor}$ (as the supervisor's input symbol) may be disallowed in the supervisor's current state. Similarly, for the actuator attacker $\attacker_a$, at some time, its input symbol $\aoutput_{\supervisor}$ may be disallowed in its current state, or its output symbol $\ainput_{\plant}$ (as the plant's input symbol) may be disallowed in the plant's current state. In all these cases, the attackers are viewed as failed since the supervisor or plant can detect the abnormality, immediately stop the supervisory control loop, and raise the alarm.

Given the discussion above, the supervisor $\supervisor$ aims to restrict the executions of the plant $\plant$ to ones that only yield words in the desired language $\desired \subseteq L(\plant)$. We denote by $L(\plant \vert  \supervisor_c)$ the language of the plant and corrupted supervisor's loop composition. This is the set of possible executions of the plant in the supervisory control loop with the attackers. Formally, the control goal is given below.

\begin{definition}
\label{def:controllability}
Let $\plant$, $\attacker_s$, and $\attacker_a$ be observable FSTs modeling the plant, sensor attacker, and actuator attacker, respectively, and $\desired \subseteq L(\plant)$ be the prefix-closed regular language capturing its desired executions. The supervisor $\supervisor$ controls the plant $\plant$ to the language $\desired$ iff the set of plant executions under the supervisor's regulation satisfies
\[
L(\plant \vert \supervisor_c) = \desired.
\]
\end{definition}
This ensures that 1) the plant execution is restricted by the desired set $\desired$ under the attacks and 2) each word in $\desired$ corresponds to a possible plant execution under certain behavior of the supervisor and attackers.

\section{Advantages of FSTs to Model Attacks}
\label{sec:attack}

This section introduces the advantages of using FSTs to model attack behaviors in supervisory control, following the discussions in Section \ref{sub:formulation-3}. Since the attack behaviors are mathematical by regular relations, we refer to the class of attacks modeled by FSTs as  \emph{regular-rewriting attacks}. The regular relations can mathematically capture the attackers' nondeterministic behaviors. These include previously studied history-independent attacks (e.g., insert, delete, and replace) \cite{wakaiki_SupervisoryControlDiscreteevent_2017,su_SupervisorSynthesisThwart_2018,goes_StealthyDeceptionAttacks_2017,carvalho_DetectionMitigationClasses_2018} and more sophisticated history-dependent attacks such as finite-memory replay attacks \cite{stuxnet,stuxnet2}, which can potentially bypass existing intrusion detectors by recording and replaying previous system executions. The work in~\cite{meiragoes2020synthesis} also implements history-dependent attacks. However, they use sensor attackers modeled by a function: the attack is determined by the plant's transition history. Comparatively, our attacker model is described by an FST allowing relatively straightforward analysis and intuitive visual representations of the attackers.


\subsection{Modeling common attacks by FSTs}
\label{sub:examples}

FSTs can model a diverse range of nondeterministic attacks in supervisory control. FST models can be derived from the regular relations that the attacks obey, e.g., by analyzing the security services and (possible) attack surfaces from the system's architecture and deployment. In return, the resilience analysis given later in this paper can guide improving the security services in the deployed system. Below, we give a few examples of modeling common attacks with FSTs.

\begin{figure*}[!t]
\centering
\subcaptionbox{Projection\label{ex:projection}}[.2\linewidth]{
\begin{tikzpicture}
  \node[state,initial,initial text=] (s) {};
  \path[->] (s) edge[loop below] node {($\ainput , \epsilon$) for $\ainput \in \ainputs \backslash \ainputs'$} ()
        (s) edge[loop above] node {($\ainput , \ainput$) for $\ainput \in \ainputs'$} ();
\end{tikzpicture}
}
\subcaptionbox{Deletion\label{ex:deletion}}[.2\linewidth]{
\begin{tikzpicture}
  \node[state,initial,initial text=] (s) {};
  \path[->] (s) edge[loop above] node {($\ainput , \epsilon)$ for $\ainput \in \ainputs \backslash \ainputs'$} ()
        (s) edge[loop below] node {$(\ainput , \ainput)$ for $\ainput \in \ainputs$} ();
\end{tikzpicture}
}
\subcaptionbox{Injection\label{ex:injection}}[.2\linewidth]{
\begin{tikzpicture}
  \node[state,initial,initial text=] (s) {};
  \path[->] (s) edge[loop above] node {($\epsilon , \ainput')$ for $\ainput' \in \ainputs'$} ()
        (s) edge[loop below] node {$(\ainput , \ainput)$ for $\ainput \in \ainputs$} ();
\end{tikzpicture}
}
\subcaptionbox{%
Replacement-Removal\label{ex:replacement-removal}}[.22\linewidth]{
\begin{tikzpicture}
  \node[state,initial,initial text=] (s) {};
  \path[->] (s) edge[loop below] node {$(\ainput , \phi(\ainput) )$ for $\ainput \in \ainputs$} ();
\end{tikzpicture}
}
\subcaptionbox{%
Injection-Removal\label{ex:injection-removal}}[.22\linewidth]{
\begin{tikzpicture}
  \node[state, initial right, initial text=] (s) {};
  \path[->]
  (s) edge[loop left] node[above,xshift=-0.6cm] {$(\ainput , \epsilon)$ for $\ainput \in \ainputs'$} ()
  (s) edge[loop above] node {$(\epsilon , \ainput)$ for $\ainput \in \ainputs'$} ()
  (s) edge[loop below] node {$(\ainput , \ainput)$ for $\ainput \in \ainputs \backslash \ainputs'$} ();
\end{tikzpicture}
}
\hspace{15pt}
\subcaptionbox{Replay Attack\label{ex:replay}}[.42\linewidth]{
\begin{tikzpicture}
  \node[state,initial,] (s) {};
  \node[state, right of=s, yshift=0.5cm, xshift=0.25cm] (s1) {};
  \node[state, right of=s, yshift=-0.5cm, xshift=0.25cm] (s2) {};
  \node[state, right of=s1, xshift=1.25cm] (s11) {};
  \node[state, right of=s2, xshift=1.25cm] (s22) {};
  \node[state, right of=s1, yshift=1.1cm, xshift=1.25cm] (s3) {};
  \node[state, right of=s2, yshift=-1.1cm, xshift=1.25cm] (s4) {};

  \path[->] (s) edge[bend left]  node[above] {$(\ainput_1 , \ainput_1)$} (s1);
  \path[->] (s) edge[bend right] node[below] {$(\ainput_2 , \ainput_2)$} (s2);
  \path[->] (s1) edge[loop above] node[above] {$(\ainput_1 \cup \ainput_2 , \ainput_1)$} ();
  \path[->] (s2) edge[loop below] node[below] {$(\ainput_1 \cup \ainput_2 , \ainput_2)$} ();
  \path[->] (s1) edge[bend left]  node[above] {$(\ainput_1 , \ainput_1)$} (s3);
  \path[->] (s2) edge[bend right] node[below] {$(\ainput_2 , \ainput_2)$} (s4);
  \path[->] (s1) edge node[above] {$(\ainput_2 , \ainput_2)$} (s11);
  \path[->] (s2) edge node[below] {$(\ainput_1 , \ainput_1)$} (s22);
  \path[->] (s3) edge[loop right] node[right] {$(\ainput_1 \cup \ainput_2 , \ainput_1)$} ();
  \path[->] (s4) edge[loop right] node[right] {$(\ainput_1 \cup \ainput_2 , \ainput_2)$} ();
  \path[->] (s11) edge[bend left] node[right] {$(\ainput_1 \cup \ainput_2 , \ainput_1)$} (s22);
  \path[->] (s22) edge[bend left] node[left] {$(\ainput_1 \cup \ainput_2 , \ainput_2)$} (s11);
\end{tikzpicture}
}
\caption{FST models for common attacks\label{fig:FST_realization}.}
\end{figure*}

\begin{example}
Let $\ainputs' \subseteq \ainputs$. The projection attack
\begin{equation} \label{eq:projection}
    \mathrm{Project}_{\ainputs'}(\ainput) = \begin{cases}
    \ainput, &\text{ if }\quad \ainput \in \ainputs' \\
    \epsilon, &\text{ otherwise,}
  \end{cases}
\end{equation}
captures the attack that results in removing all symbols that belong to $\ainputs\setminus\ainputs'$.
On the other hand, the nondeterministic deletion attack is defined as
\begin{equation} \label{eq:deletion}
  \mathrm{Delete}_{\ainputs'}(\ainput) = \begin{cases}
    \ainput, &\text{ if }\quad\ainput \in \ainputs' \\
    \epsilon \text{ or } i, &\text{ otherwise.}
  \end{cases}
\end{equation}
It extends the $\mathrm{Project}_{\ainputs'}$ attack as
it captures that
the attacker may (or may not) remove symbols from $\ainputs\setminus\ainputs'$; e.g., if $\ainputs' = \ainputs$ this model can be used to capture Denial-of-Service attacks~\cite{wood2002denial} over the communication network. Finally, the nondeterministic injection attack is defined as
\begin{equation} \label{eq:injection}
  \mathrm{Inject}_{\ainputs'}(\varepsilon) = \ainput \text{ where } \ainput \in \ainputs'. 
\end{equation}
In it, a finite number of symbols from $\ainputs'$ can be added before and/or after the symbols. These attacks 
can be represented by FSTs as shown in Fig.~\ref{ex:projection}, Fig.~\ref{ex:deletion} and Fig.~\ref{ex:injection}, respectively.
\end{example}

\begin{example}
\label{ex:rep-rem}
A replacement-removal attack defined by the rule $\phi: \ainputs \rightarrow 2^{\ainputs \cup \epsilon}$ is represented by an FST as shown in Fig.~\ref{ex:replacement-removal}.
\end{example}

\begin{example}
\label{ex:ins-rem}
Let $\ainputs' \subseteq \ainputs$. An injection-removal attack nondeterministically injects or removes symbols in $\ainputs'$ from a word. This is modeled by the FST in Fig.~\ref{ex:injection-removal}.
\end{example}

Beyond the simple attackers in Example~3-5, FSTs can also model complex history-dependent attacks, motivated by cyber/network attacks implemented by embedded programs (e.g., malware). Since the attackers modeled by FSTs have (internal) states, they can perform different attack actions depending on their current state, which is affected by their previous attack actions. Take the replay attack as an example, which is a common strategy for launching cyber attacks~\cite{miao_cdc13, mo_procIEEE12}. It works by first recording a fragment of symbols and then replaying it repeatedly to trap the system. Implementing this attack requires the attacker to know when to continue recording or start replaying.

\begin{example}
A replay attack records a prefix of a word and replaces the rest with the repetitions of the recorded prefix, with the prefix size being bounded by the finite-memory capacity (i.e., size) $N$. This relation is regular, so can be modeled by FSTs. For example, a replay attack recording a prefix of length up to $N = 2$ for any word $\ainputs=\{\ainput_1, \ainput_2\}$ can be modeled by an FST as shown in Fig.~\ref{ex:replay}. Note that the FST can be viewed as the parallel composition of two replay attacks recording prefixes of length $1$ and $2$, respectively.
\end{example}

In~\cite{meiragoes2020synthesis}, the attackers are also modeled with states. However, these states are shared with the plant (referred to as the environment). For different plant models, this approach requires building a new combined model to capture the same replay attack. On the other hand, our approach allows modeling a replay attack separately and then composing it with the plant model to study its impact. Our compositional approach can help the users to 1) model each component separately, 2) study the impact of different attacks on the same plant, 3) study the impact of multiple combined attacks.

\subsection{Composition of multiple attacks}
\label{sub:composition}

Multiple attacks modeled by FSTs can
be combined into an overall FST model. This justifies our modeling from the general system architecture in Fig.~\ref{fig:diagram} to the control diagram in Fig.~\ref{fig:control}. Specifically, in Fig.~\ref{fig:diagram}, there may be coordinated attacks from multiple deployed attack vectors with different ``point-of-entries'' from the sensors, actuators, and communication networks. 
For example, on the sensor side, there may be false data injection via sensor spoofing~\cite{gps_spoof1,shepard12} together with denial-of-service (DoS) attacks on network transmissions. The overall effect of those attacks is equivalent to their serial composition as illustrated in Fig.~\ref{fig:serial}, which is represented by a single FST $\attacker_s$ in Fig.~\ref{fig:control}. 

Similarly, if there is one of several possible attacks that may be deployed at the same ``point-of-entry'' as studied in~\cite{wakaiki_SupervisoryControlDiscreteevent_2017}, the overall effect, and the corresponding FST model, is equivalent to the parallel composition of the two FSTs capturing the basic attacks, as shown in Fig.~\ref{fig:parallel}. Algorithmically, the parallel composition is computed as follows.
\begin{enumerate}
  \item Compute the union of the states, final states, and transitions.
  \item Add a new starting state with transitions $\epsilon/\epsilon$ to the initial states $\ainit$ and $\ainit'$.
\end{enumerate}


\begin{figure}[!t]
\centering
\subcaptionbox{{When $n$ attacks happen simultaneously (e.g., false data injection attack on sensors $\attacker_1$, denial-of-service attack on network $\attacker_2$, ...), the overall effect can be captured by the serial composition.}\label{fig:serial}}[.45\linewidth]{
\begin{tikzpicture}
 \node (a0) {};
 \node [draw, right of=a0]  (a1) {$\attacker_1$};
 \node [right of=a1]  (a2) {$\ldots$};
 \node [draw, right of=a2]  (a3) {$\attacker_n$};
 \node [right of=a3] (a4) {};
 \draw [->, >=latex] (a0) -- (a1);
 \draw [->, >=latex] (a1) -- (a2);
 \draw [->, >=latex] (a2) -- (a3);
 \draw [->, >=latex] (a3) -- (a4);
\end{tikzpicture}
}
\hspace{0.3cm}
\subcaptionbox{{When (only) one of $n$ possible attacks may occur at a ``point-of-entry'' (i.e., attack point), the overall effect can be captured by the parallel composition of the attack models.}\label{fig:parallel}}[.45\linewidth]{
\begin{tikzpicture}
 \node (a0) {};
 \node [right of=a0]  (a2) {$\ldots$};
 \node [draw, above of=a2, yshift=-0.4cm]  (a1) {$\attacker_1$};
 \node [draw, below of=a2, yshift=0.4cm]  (a3) {$\attacker_n$};
 \node [right of=a2] (a4) {};
 \draw [->, >=latex, dashed] (a0.center) -- (a2);
 \draw [->, >=latex, dashed] (a2) -- (a4.center);
 \draw [->, >=latex] (a0.center) |- (a1);
 \draw [->, >=latex] (a1) -| (a4.center);
 \draw [->, >=latex, dashed] (a0.center) |- (a3);
 \draw [->, >=latex, dashed] (a3) -| (a4.center);
\end{tikzpicture}
}

\par\bigskip

\subcaptionbox{Modeling frequency constraints\label{fig:frequency}; e.g., the operational constraint that an FST-based projection attacker $\attacker$ happens once every three steps can be captured by augmenting the FST model}[.8\linewidth]{
\begin{tikzpicture}
\node[initial, state, initial text=] (0) {0};
\node[state, right of = 0, xshift = 0.8cm] (1) {1};
\node[state, below of = 1, yshift = -0.8cm] (3) {$2$};

\path[->] (0) edge[bend left] node[above] {$(\ainput , \ainput)$} (1);
\path[->] (1) edge[bend left] node[right] {$(\ainput , \ainput)$} (3);
\path[->] (3) edge[bend left] node[left] {$(\ainput , \epsilon)$} (0) ;
\path[->] (3) edge[bend right] node[left] {$(\ainput , \ainput)$} (0) ;





\end{tikzpicture}
}
\caption{Modeling attack composition and attack constraints\label{fig:fst_compose}.}
\end{figure}

\subsection{Imposing constraints on attacks}
\label{sub:constraints}

FSTs also facilitate imposing (operational) constraints and transform history-independent attacks (e.g., injection attacks from~\eqref{eq:injection}) to history-dependent attacks. 
For example, in some networked control systems, cryptographic primitives (e.g., Message Authentication Codes) can only be intermittently used due to resource constraints, and thus can only intermittently prevent injection attacks~\cite{jovanov_tac19,lesi_tecs17,lesi_tcps20}. FSTs can facilitate modeling such constraints as frequency constraints on the attacker as studied in~\cite{su_SupervisorSynthesisThwart_2018}.

To illustrate this, we take the simple FST in Figure~\ref{ex:projection} and show how to impose a frequency constraint.

\begin{example}
The FST in Figure~\ref{ex:projection} can remove the input symbol $\ainput$ non-deterministically. The removal can happen for all symbols in the worst case. Suppose we want to constrain the frequency to at most once every three symbols. Then, we can model it by the FST model in Figure~\ref{fig:frequency} which is modified from the former FST.
\end{example}

\section{Attack-Resilient Supervisor Design}
\label{sec:design}

This section presents theories and algorithms to synthesize resilient supervisors for the formulation in Definition \ref{def:controllability}. 
Since the plants are modeled by FSTs and can generate different input and output symbols, we consider supervisors that can monitor a pair of different input and output symbols each time, as discussed in Section \ref{sub:formulation-1}. Synthesizing the resilient supervisor that satisfies Definition \ref{def:controllability} is not always feasible. Thus, we consider a related problem of designing a candidate resilient supervisor that can restrict $L(\plant \vert \supervisor_c)$ where $\supervisor_c = \attacker_a \circ \supervisor \circ \attacker_s$ to the minimal feasible super-language of $\desired$, i.e., finding a $\supervisor$ such that $L(\plant \vert \supervisor_c)$ is the smallest possible language that contains $\desired$. 


\begin{definition}
\label{def:weak supervisor}
The FST $\supervisor$ is a candidate resilient supervisor for a desired language $\desired$ iff
\begin{enumerate}
\item $\desired \subseteq L(\plant  \vert  \attacker_a \circ \supervisor \circ \attacker_s)$, and
\item for any $\supervisor'$ that satisfies 
$\desired \subseteq L(\plant  \vert  \attacker_a \circ \supervisor' \circ \attacker_s)$, it holds that 
$L(\plant  \vert  \attacker_a \circ \supervisor \circ \attacker_s) \subseteq L(\plant  \vert  \attacker_a \circ \supervisor' \circ \attacker_s)$.
\end{enumerate}
\end{definition}

If $ L(\plant  \vert  \attacker_a \circ \supervisor \circ \attacker_s) $ is equal to $\desired$, then the candidate resilient supervisor is the resilient supervisor that satisfies Definition \ref{def:controllability}. In the following, we study attacks on sensors in Section \ref{sub:sensor attacks}, on actuators in Section \ref{sub:actuator attacks}, and on both sensors and actuators in Section \ref{sub:both attacks}.

\subsection{Design resilient supervisors to sensor attacks}
\label{sub:sensor attacks}

To synthesize the candidate resilient supervisor in Definition \ref{def:weak supervisor} with only the sensor attacker $\attacker_s$, we use the FST inverse $\attacker_s^{-1}$ as the countermeasure.

\begin{definition}
\label{def:fst_inverse}
Given an FST, $\attacker = (\astates, \ainit, \ainputs, \aoutputs, \atransitions, \allowbreak \afinal)$, the FST inverse is also an FST denoted by $\attacker^{-1} = (\astates, \ainit, \aoutputs, \ainputs, \atransitions^{-1}, \afinal)$, where the transition $(\astate, \aoutput, \ainput, \astate') \in \atransitions^{-1}$ if and only if $(\astate, \ainput, \aoutput, \astate') \in \atransitions$.
\end{definition}

The inverse of an FST is derived by flipping the input/output symbols on each transition, as given in definition~\ref{def:fst_inverse}. By inverting an FST, the regular relation it defines is also reverted, as stated by the following lemma \cite{holcombe_AlgebraicAutomataTheory_1982}, and definition~\ref{def:fst_inverse} provides an FST realization for the inverse of regular relations.

\begin{lemma}
\label{lem:inverse}
For an FST $\attacker$,
it holds that $\rr_{\attacker^{-1}} = \rr_{\attacker}^{-1}$,
where $\rr_\attacker$ denotes the regular relation defined by $\attacker$.
\end{lemma}







The composition of a relation and its inverse includes the identity map. Thus, we have the following lemma on the composition of an FST and its inverse.

\begin{lemma}
\label{lem:id}
For any FST $\attacker$, it holds that 
\begin{align*}
& \forall I \in L_{\mathrm{in}}(\attacker). \ I \in \rr_{\attacker \c \attacker^{-1}} (I),   
\\ & \forall I \in L_{\mathrm{out}}(\attacker). \ I \in \rr_{\attacker^{-1} \c \attacker} (I).
\end{align*}
Equivalently, it holds that
\[
\rr_{\mathcal{M}_{L_{\mathrm{in}} (\attacker)}} \subseteq \rr_{\attacker \c \attacker^{-1}}, \quad \rr_{\mathcal{M}_{L_{\mathrm{out}} (\attacker)}} \subseteq \rr_{\attacker^{-1} \c \attacker}.
\]
Here, $L_{\mathrm{in}} (\attacker)$ and $L_{\mathrm{out}} (\attacker)$ are the input and output languages of $\attacker$, and $\mathcal{M}_L$ is the automaton realization (thus also an FST) of the regular language $L$. 
\end{lemma}

Recall Theorem \ref{thm:1}. For the desired language $\desired \subseteq L(\plant)$ of the plant $\plant$, the FST $\supervisor$ can supervise the plant's language in the closed loop without attacks, i.e., let $L(\plant \vert \supervisor) = \desired$, if $\supervisor= \mathcal{M}_\desired^{-1}$ where $\mathcal{M}_\desired$ is a deterministic FST realizing $\desired$. The supervisor should be the inverse of $\mathcal{M}_\desired$ since the plant and supervisor have flipped input and output. Now, after introducing the sensor attacker $\attacker_s$, we show that the composition of the inverse of $\attacker_s$ and $\mathcal{M}_\desired^{-1}$ can potentially serve as a resilient supervisor, as stated below. Intuitively, in the supervisor $\supervisor = \attacker_s^{-1} \c \mathcal{M}_\desired^{-1}$, the part $\attacker_s^{-1}$ partially reverts the attacks of $\attacker_s$ by finding out the possible true words before the attacks, and then the part $\mathcal{M}_\desired^{-1}$ filters out undesirable words. Thus, the natural choice of the candidate supervisor is $(\mathcal{M}_\desired \c \attacker_s)^{-1} = \attacker_s^{-1} \c \mathcal{M}_\desired^{-1}$. The theorem below formally explains this approach.


\begin{theorem}
\label{prop:sensor attack}
With only the sensor attacker $\attacker_s$, for any FST plant $\plant$ and desired prefix-closed regular language $\desired \subseteq L(\plant)$, the deterministic FST satisfying
$$\rr_{\supervisor_{\min}} = \rr_{\attacker_s^{-1} \c \mathcal{M}_\desired^{-1}}$$ 
is a candidate resilient supervisor. 
It is a resilient supervisor iff 
$$L(\mathcal{M}_\desired \c \attacker_s \c \attacker_s^{-1}) \cap L(\plant) = \desired.$$
\end{theorem}

\begin{proof}
We defer the proof until Theorem \ref{thm:both attack}.
\end{proof}

\subsection{Design resilient supervisors to actuator attacks}
\label{sub:actuator attacks}




Similar to \ref{sub:sensor attacks}, when there is only the actuator attacker $\attacker_a$, we use the FST inverse $\attacker_a^{-1}$ as the countermeasure. The theorem below formally explains this approach.

\begin{theorem}
\label{prop:actuator attack}
With only the actuator attacker $\attacker_a$, for any plant $\plant$ and desired prefix-closed regular language $\desired \subseteq L(\plant)$, the deterministic FST satisfying
$$\rr_{\supervisor_{\min}} = \rr_{\mathcal{M}_\desired^{-1} \c \attacker_a^{-1}}$$  
is a candidate resilient supervisor. It is a resilient supervisor iff
$$L(\attacker_a^{-1} \c \attacker_a \c \mathcal{M}_\desired) \cap L(\plant) = \desired.$$
\end{theorem}

\begin{proof}
We defer the proof until Theorem \ref{thm:both attack}.
\end{proof}

Intuitively, the supervisor is corrupted by the composition with $\attacker_a$, thus the natural choice of the candidate supervisor is $ \mathcal{M}_\desired^{-1} \c \attacker_a^{-1}$, according to Theorem \ref{thm:1}.

\subsection{Design resilient supervisors to both sensor and actuator attacks}
\label{sub:both attacks}

When both the sensor and actuator attackers $\attacker_s$ and $\attacker_a$ exist, we combine the approaches from Section \ref{sub:sensor attacks} and \ref{sub:actuator attacks} and use the FST inverse $\attacker_s^{-1}$ and $\attacker_a^{-1}$ as the countermeasure. Again, we can think of the composition $\attacker_a \c \plant \c \attacker_s$ as the corrupted plant, and $L(\attacker_a \c \mathcal{M}_\desired \c \attacker_s)$, thus the natural choice of the candidate supervisor is $(\attacker_a \c \mathcal{M}_\desired \c \attacker_s)^{-1} = \attacker_s^{-1} \c \mathcal{M}_\desired^{-1} \c \attacker_a^{-1}$, according to Theorem \ref{thm:1}. The theorem below formally explains this approach.


\begin{theorem}
\label{thm:both attack}
For any plant $\plant$ and desired prefix-closed regular language $\desired \subseteq L(\plant)$, the deterministic FST satisfying
\begin{equation}
\label{eq:thm-both-attack-1}
\rr_\supervisor = \rr_{\attacker_s^{-1} \c \mathcal{M}_\desired^{-1} \c \attacker_a^{-1}}
\end{equation}  
is a candidate resilient supervisor. It is a resilient supervisor iff 
\begin{equation}
\label{eq:thm-both-attack-2}
L(\attacker_a^{-1} \c \attacker_a \c \mathcal{M}_\desired \c \attacker_s \c \attacker_s^{-1}) \cap L(\plant) = \desired.
\end{equation}  
\end{theorem}

\begin{proof}
We start with proving the deterministic FST $\supervisor_{\min}$ satisfying \eqref{eq:thm-both-attack-1} is a candidate resilient supervisor. The existence of $\supervisor_{\min}$ results from Lemma \ref{lem:1} and that $\rr_{\attacker_s^{-1} \c \mathcal{M}_\desired^{-1} \c \attacker_s^{-1}}$ is a regular relation.

First, we prove the supervisor $\supervisor_{\min}$ satisfies Condition 2) of Definition \ref{def:weak supervisor}. Consider another supervisor $\supervisor$ that satisfies Condition 1) of Definition \ref{def:weak supervisor}. Thus, it should allow any word in $\desired$ for the plant under the attackers. Namely, any pair of input and output words $I_\plant = \ainput_1 \ldots \ainput_n$ and $O_\plant = \aoutput_1 \ldots \aoutput_n$ satisfying $(\ainput_1, \aoutput_1) \ldots (\ainput_n, \aoutput_n) \in \desired$, i.e., 
\begin{equation} \label{eq:pf-1}
(I_\plant, O_\plant) \in \rr_{\mathcal{M}_\desired},
\end{equation} 
should be allowed for the plant. Here, $\ainput_1, \aoutput_1, \ldots \ainput_n, \aoutput_n$ may be the empty symbol $\varepsilon$. Similarly, the pair of input and output words $O_\supervisor$ and $I_\plant$, should be allowed by the attacker $\attacker_a$ if 
\begin{equation} \label{eq:pf-2}
(O_\supervisor, I_\plant) \in \rr_{\attacker_a}.
\end{equation} 
And the pair of input and output words $O_\plant$ and $I_\supervisor$, should be allowed by the attacker $\attacker_a$ if 
\begin{equation} \label{eq:pf-3}
(O_\plant, I_\supervisor) \in \rr_{\attacker_s}.
\end{equation} 
Since the plant $\plant$ and attackers $\attacker_s$ and $\attacker_a$ are deterministic, their executions corresponding determined by $I_\plant, O_\plant, I_\supervisor, O_\supervisor$. Combining \eqref{eq:pf-1}\eqref{eq:pf-2}\eqref{eq:pf-3} gives
$$(O_\supervisor, I_\supervisor) \in \rr_{\attacker_a} \c \rr_{\mathcal{M}_\desired} \c \rr_{\attacker_s}.
$$
That is, by Lemma \ref{lem:fst-composition} and \ref{lem:inverse},
\begin{equation} \label{eq:pf-4}
(I_\supervisor, O_\supervisor) \in \rr_{\attacker_s^{-1} \c \mathcal{M}_\desired^{-1} \c \attacker_s^{-1}}.
\end{equation}
By construction, any such pair of $I_\supervisor$ and $O_\supervisor$ should be accepted by $\supervisor$, thus 
$$\rr_{\attacker_s^{-1} \c \mathcal{M}_\desired^{-1} \c \attacker_s^{-1}} = \rr_{\supervisor_{\min}} \subseteq \rr_{\supervisor}.
$$
Therefore, $\supervisor_{\min}$ is the ``minimal'' supervisor.

Second, we prove the supervisor $\supervisor_{\min}$ satisfy Condition 1) of Definition \ref{def:weak supervisor}. Again, consider a pair of words $I_\plant$ and $O_\plant$ satisfying \eqref{eq:pf-1} generated from the plant $\plant$. Let $I_\supervisor$ and $O_\supervisor$ be the corresponding corrupted words sent to the supervisor. Then, the pair $(O_\supervisor, I_\plant)$ should satisfies \eqref{eq:pf-2} and  the pair $(O_\plant, I_\supervisor)$ should satisfies \eqref{eq:pf-3}. Since the supervisor $\supervisor_{\min}$ and attackers $\attacker_s$ and $\attacker_a$ are deterministic, their executions corresponding determined by $I_\plant, O_\plant, I_\supervisor, O_\supervisor$. Combining the above conditions, we have that any such pair of $I_\supervisor$ and $O_\supervisor$ should satisfy \eqref{eq:pf-4}, and thus allowed by the supervisor $\supervisor_{\min}$.

Finally, we show that the left-hand side of \eqref{eq:thm-both-attack-2} gives the language $L(\plant \ \vert \ \attacker_a \circ \supervisor \circ \attacker_s)$ of the plant in the closed loop with supervisor and attackers. Following the formulation in Section \ref{sub:formulation-3}, the set of plant words (i.e., sequences of input/output symbol pairs of the plant) allowed by the supervisor $\supervisor_{\min}$ and attackers $\attacker_s$ and $\attacker_a$ is 
\begin{equation} \label{eq:pf-5}
  L((\attacker_s \c \supervisor_{\min} \c \attacker_a)^{-1}) = L(\attacker_a^{-1} \c \attacker_a \c \mathcal{M}_\desired \c \attacker_s \c \attacker_s^{-1}).
\end{equation}
This result can be derived by viewing $\attacker_s \c \supervisor_{\min} \c \attacker_a$ as a corrupted supervisor and applying Theorem~\ref{thm:1}. The inverse is because the plant's input is the supervisor's output and the plant's output is the supervisor's input. Therefore, the plant language $L(\plant \ \vert \  \attacker_a \circ \supervisor \circ \attacker_s)$ is the intersection of \eqref{eq:pf-5} with $L(\supervisor)$. When~\eqref{eq:thm-both-attack-2} holds, we have $L(\plant \ \vert \ \attacker_a \circ \supervisor \circ \attacker_s) = \desired$, thus $\supervisor_{\min}$ is a resilient supervisor.
\end{proof}

Theorem~\ref{thm:both attack} gives Algorithm~\ref{alg:both attack} for resilient supervisors to the actuator and sensor attacks. This algorithm computes a minimal superlanguage of $\desired$. If the aforementioned super language equates to $\desired$, then the designed supervisor is resilient. The algorithm only involves FST inverse and composition and thus has polynomial complexity in time. Our previous work \cite{wang2019attack} develops a method to compute the largest subset of $\desired$ that the plant can follow with supervisory control under the influence of attackers. The latter is known as a maximal sublanguage. However, this approach is much more computationally expensive than the former since it relies on fixed point operations. Below is an illustrative example. A more complex example is given in Section \ref{sec:sim}.

\begin{algorithm}[!t]
\caption{Resilient supervisor for both sensor and actuator attacks}
\label{alg:both attack}
\begin{algorithmic}[1]
\Require{Desired language $\desired \subseteq L(\plant)$, and attackers $\attacker_s$ and $\attacker_a$.}
\State{Build an FST $\mathcal{M}_\desired$ realizing $\desired$.}
\State{Compute inverse $\attacker_a^{-1}$, $\mathcal{M}_\desired^{-1}$ and $\attacker_s^{-1}$.}
\State{Compute composition $\supervisor = \attacker_s^{-1} \c \mathcal{M}_\desired^{-1} \c \attacker_a^{-1}$.}
\If{$L((\attacker_s \c \supervisor \c \attacker_a)^{-1}) \cap L(\plant) = \desired$}
\State \Return $\supervisor$
\Else
\State \Return Infeasible
\EndIf
\end{algorithmic}
\end{algorithm}

\begin{example}
\label{ex:both_attack}
As shown in Fig.~\ref{fig:both_example}, consider 
\begin{itemize}
\item  set of input and output symbols $\ainputs = \{\ainput_1,\ainput_2, \ainput_3,\ainput_4, \ainput_5\}$ and $\aoutputs = \{\aoutput_1,\aoutput_2, \aoutput_3,\aoutput_4, \aoutput_5\}$,
\item  a plant $\plant$ with language $\overline{\{t_1t_2, t_3t_4, t_5\}^*}$ as shown and defined in Fig.~\ref{fig:both_plant},
\item  a desired language $\desired = \overline{ \{ t_1t_2\}^*}$ realized by FST $\mathcal{M}_\desired$ in Fig.~\ref{fig:both_desired},
\item two actuator attackers: $\attacker_{a1}$ from Fig.~\ref{fig:both_actuator_1}  rewrites $\aoutput_1$ to $\aoutput_3$ and vice versa while $\attacker_{a2}$ from Fig.~\ref{fig:both_actuator_2} rewrites $\aoutput_3$ to $\aoutput_1$ only,
\item two sensor attackers: $\attacker_{s1}$ from Fig.~\ref{fig:both_sensor_1}  rewrites $\ainput_5$ to $\ainput_1$ and vice versa while $\attacker_{s2}$ from Fig.~\ref{fig:both_sensor_2} rewrites $\ainput_1$ to $\ainput_5$ only.
\end{itemize}
In this example, we have shown two different types of attacks and supervisors for the same plant FST and desired language, $\desired$. The supervisor shown in Fig.~\ref{fig:both_supervisor_1} was designed to counter $\attacker_{a1}$ and $\attacker_{s1}$. Similarly, the supervisor in Fig.~\ref{fig:both_supervisor_2} was designed to counter $\attacker_{a2}$ and $\attacker_{s2}$.

Both supervisors were designed using Algorithm 3. The supervisor language $L(\attacker_s^{-1} \c \mathcal{M}_\desired^{-1} \c \attacker_s^{-1})$ was computed to determine if the plant can be restricted to $\desired$. The choice of $\attacker_{a2}$ and $\attacker_{s2}$ does not allow for a feasible resilient supervisor design because the $\attacker_{s2}$ forces $\ainput_5$ to be the only symbol sent to the plant. The supervisor will then receive $\aoutput_5$ and raise the alarm. Thus, no transitions will ever occur. 
\end{example}

In practice, it is unrealistic to assume a design engineer has complete knowledge of the attackers on a CPS. However, this algorithm is still very useful because it can quickly test for resilient supervisor existence based on a wide variety of nondeterministic attacks. In practice, our proposed FST attack model would act as a conservative overestimate of the true attacker.

\begin{figure}
\centering
\subcaptionbox{Plant.\label{fig:both_plant}}[.45\linewidth]{
\begin{tikzpicture}
  \node[state,initial, initial text=] (s) {$0$};
  \node[state,right of=0, xshift = 1.00cm] (1) {$1$};
  \node[state,below of=0, xshift = -0.4cm, yshift = -1.00cm] (2) {$2$};
  \path[->] (0) edge[bend left] node[above] {$t_1: (\ainput_1, \aoutput_1)$} (1); 
  \path[->]  (1) edge[bend left] node[below] {$t_2:(\ainput_2, \aoutput_2)$} (0);
  \path[->]  (0) edge[bend left] node[right] {$t_3:(\ainput_1, \aoutput_3)$} (2);
  \path[->]  (2) edge[bend left] node[left] {$t_4:(\ainput_4, \aoutput_4)$} (0);
  \path[->]  (0) edge[loop above] node[above] {$t_5:(\ainput_5, \aoutput_5)$} ();
\end{tikzpicture}
}
\subcaptionbox{FST of $\desired$.\label{fig:both_desired}}[.5\linewidth]{
\begin{tikzpicture}
  \node[state,initial, initial text=] (0) {$0$};
  \node[state,right of=0] (1) {$1$};

  \path[->] (0) edge[bend left] node[above] {$t_1:(\ainput_1, \aoutput_1)$} (1);
  \path[->] (1) edge[bend left] node[below] {$t_2:(\ainput_2, \aoutput_2)$} (0);

\end{tikzpicture}
}

\par\bigskip
\subcaptionbox{Actuator Attacker I.\label{fig:both_actuator_1}}[.45\linewidth]{
\begin{tikzpicture}
  \node[state,initial, initial text=] (0) {$0$};
  \path[->] (0) edge[loop above] node[above] {$(\aoutput_3 , \aoutput_1)$} (0);
  \path[->] (0) edge[loop right] node[right] {$(\aoutput_1 , \aoutput_3)$} (0);
  \path[->] (0) edge[loop below] node[below] {$(\aoutput_i , \aoutput_i),  \forall \aoutput_i \neq \aoutput_1, \aoutput_3$} (0);
\end{tikzpicture}
}
\subcaptionbox{Actuator Attacker II.\label{fig:both_actuator_2}}[.45\linewidth]{
\begin{tikzpicture}
  \node[state,initial, initial text=] (0) {$0$};
  \path[->] (0) edge[loop above] node[above] {$(\aoutput_3 , \aoutput_1)$} (0);

  \path[->] (0) edge[loop below] node[below] {$(\aoutput_i , \aoutput_i),  \forall \aoutput_i \neq \aoutput_3$} (0);
\end{tikzpicture}
}

\par\bigskip
\subcaptionbox{Sensor Attacker I\label{fig:both_sensor_1}}[.45\linewidth]{
\begin{tikzpicture}
  \node[state,initial, initial text=] (0) {$0$};
  \path[->] (0) edge[loop above] node[above] {$(\ainput_5 , \ainput_1)$} (0);
  \path[->] (0) edge[loop right] node[right] {$(\ainput_1 , \ainput_5)$} (0);
  \path[->] (0) edge[loop below] node[below] {$(\ainput_i , \ainput_i),  \forall \ainput_i \neq \ainput_1, \ainput_5$} (0);
\end{tikzpicture}
}
\subcaptionbox{Sensor Attacker II\label{fig:both_sensor_2}}[.45\linewidth]{
\begin{tikzpicture}
  \node[state,initial, initial text=] (0) {$0$};
  \path[->] (0) edge[loop above] node[above] {$(\ainput_1 , \ainput_5)$} (0);
  \path[->] (0) edge[loop below] node[below] {$(\ainput_i , \ainput_i),  \forall \ainput_i \neq \ainput_1$} (0);
\end{tikzpicture}
}

\par\bigskip
\subcaptionbox{Designed Supervisor I\label{fig:both_supervisor_1}}[.45\linewidth]{
\begin{tikzpicture}
  \node[state,initial, initial text=] (0) {$000$};
  \node[state,right of=1] (1) {$010$};
  \path[->] (0) edge[bend left] node[above] {$(\aoutput_3 , \ainput_5)$} (1);
  \path[->] (1) edge[bend left] node[below] {$(\aoutput_2 , \ainput_2)$} (0);
\end{tikzpicture}
}
\subcaptionbox{Infeasible Supervisor II\label{fig:both_supervisor_2}}[.45\linewidth]{
\begin{tikzpicture}
  \node[state,initial, initial text=] (0) {$000$};
  \node[state,right of=1] (1) {$010$};
  \path[->] (0) edge[bend left] node[above] {$(\aoutput_3 , \ainput_5)$ or $(\aoutput_3 , \ainput_1)$ or $(\aoutput_1 , \ainput_5)$} (1);
  \path[->] (1) edge[bend left] node[below] {$(\aoutput_2 , \ainput_2)$} (0);
\end{tikzpicture}
}
\caption{Example supervisors for both sensor and actuator attacks\label{fig:both_example}.}
\end{figure}

\subsection{Relationship to Previous Work} \label{rem:recover}

Due to its generality, the presented FST-based framework described in Section~\ref{sec:formulation} and~\ref{sec:attack} 
can emulate several previously-studied setups of supervisory control.
For example, we can emulate the classical supervisory control~\cite{cassandras_IntroductionDiscreteEvent_2008} with uncontrollable symbols $\ainputs_{uc}$ 
in our framework by adopting the following setups in Fig.~\ref{fig:control} by letting
$\plant$ be the plant (an automaton) of interest and the actuator attacker be $\attacker_a^{(s)} = \mathrm{Inject}_{\ainputs_{uc}}$ and $\mathrm{Inject}_{\ainputs_{uc}} \circ \plant$ be the serial composition of the injection attack from~\eqref{eq:injection} and the plant. This attacker injects uncontrollable symbols in $\ainputs_{uc}$ whenever they are acceptable by the plant.
The supervisory control under sensor/actuator enablement and disablement attacks~\cite{carvalho_DetectionMitigationClasses_2018} can be emulated in our framework 
by
letting the sensor/actuator attacks be $\attacker_a^{(ed)} \circ \attacker_a^{(s)}$, where $\attacker_a^{(s)}$ is defined above and $\attacker_a^{(ed)}$ is the injection-removal attack on a set of vulnerable control symbols from Example~\ref{ex:ins-rem}.
Finally, the supervisory control under replacement-removal sensor attacks~\cite{wakaiki_SupervisoryControlDiscreteevent_2017} can be emulated in our framework 
by
letting the sensor attacks be $\attacker_s^{(rr)} \circ \attacker_s^{(s)}$, where $\attacker_s^{(s)}$ is defined above and $\attacker_s^{(rr)}$ is the replacement-removal attack from Example~\ref{ex:rep-rem}.


\section{Case Studies: ARSC Tool and Synthesis Scalability}
\label{sec:sim}

Based on the proposed algorithms, we developed an open-source tool \emph{ARSC}~\cite{arsc} based on the OpenFst library~\cite{allauzen_OpenFstGeneralEfficient_2007} and illustrated its efficiency on problems on different scales. For all evaluations, the tool was executed on an Intel Core i7-7700K CPU, and the execution time and memory usage were measured.

To illustrate the effectiveness of our approach, we consider a scheduling problem from~\cite{cassandras_IntroductionDiscreteEvent_2008}, where $n$ players independently require service for $m$ sequential tasks $t_{ij}, \ i \in [n], j \in [m]$ on a central server.
The tasks of each player have to be served in the index order.
The sensors are corrupted by an attacker that removes the tasks performed by the first player, and the actuator attacks nondeterministically rotating the input sequence $t_{1j}t_{2j}\dots t_{(n-1)j}t_{nj}$ to $t_{2j}t_{3j}\dots t_{nj}t_{1j}$ for any task index $j \in [m]$.
For $n = 2, m = 2$, the desired language $\desired$ for the system, as well as the attacks are modeled as shown in Fig.~\ref{fig:sim_example}. From Theorem~\ref{thm:both attack}, the language $\desired$ is controllable and the attacks can be countered by the supervisor constructed by Algorithm~\ref{alg:both attack}. The supervisor for the case $n=2, m=2$ is displayed in Fig.~\ref{fig:sim_supervisor}.

The complexity of the supervisor synthesis algorithms is determined by the composition operation.
The composition $\mathcal{A}_1 \circ \mathcal{A}_2$ requires $O( \abs{\astates_{\mathcal{A}_1}} \abs{\astates_{\mathcal{A}_2}} D_{\mathcal{A}_1}(\text{log}(D_{\mathcal{A}_2})+M_{\mathcal{A}_2}))$ time and $O( \abs{\astates_{\mathcal{A}_1}} \abs{\astates_{\mathcal{A}_2}} D_{\mathcal{A}_1}M_{\mathcal{A}_2})$ space where $\abs{\astates_\cdot}$, $D_\cdot$ and $M_\cdot$ denote the number of states, the maximum out-degree and the maximum multiplicity for the FST, respectively~\cite{allauzen_OpenFstGeneralEfficient_2007}.
The order in which the composition operations are performed can also change the overall complexity.
For simplicity, the term $\mathcal{P}^{-1}$ is dropped and the supervisor is computed as $ (\mathcal{A}_s^{-1} \circ \mathcal{M}_\desired^{-1}) \circ \mathcal{A}_a^{-1}$ in our implementation.
Therefore, the overall time complexity is reduced to $O( \abs{\astates_{\mathcal{M}_\desired}} \abs{\astates_{\mathcal{A}_a}} D_{\mathcal{A}_s}\text{log}(D_{\mathcal{A}_a}))$ where $ \astates_{\mathcal{M}_\desired} \sim O((m+1)^n)$ and $ \astates_{\mathcal{A}_a}{=}D_{\mathcal{A}_s}{=}D_{\mathcal{A}_a} \sim O(mn)$ for this problem.

Table~\ref{tab:sim_comp} shows the running times of the algorithm averaged over $100$ synthesis and the maximum amount of memory used during the tool execution for different values of $n$ and $m$. We can observe that a tenfold increase in the number of states in $\mathcal{M}_\desired$ increases the execution time and the memory usage by at most $100$ times. This sub-quadratic increase is a consequence of the composition operations performed by the algorithm.

\begin{table}[!t]
    \centering
    \caption{Execution time and memory usage of the supervisory synthesizing algorithm for different values of $n$ and~$m$.} \label{tab:sim_comp}
    \begin{tabular}{|c|c|c|c|c|}
        \hline
        $m$ & $n$ & $ \abs{\astates_{\mathcal{M}_\desired}} $ & Time (ms) & Memory (MB) \\
        \hline
        $9$ & $2$ & $10^2$ & $15.566$ & $0.20944$ \\
        \hline
        $9$  & $3$ & $10^3$ & $16.309$ & $2.1642$ \\
        \hline
        $99$ & $2$ & $10^4$ & $20.852$ & $47.563$ \\
        \hline
        $9$ & $5$ & $10^5$ & $99.535$ & $340.26$ \\
        \hline
        $99$ & $3$ & $10^6$ & $721.68$ & $7106.7$ \\
        \hline
    \end{tabular}
\end{table}



\begin{figure*}[!t]
\centering
\subcaptionbox{Plant \label{fig:sim_plant}}[0.15\linewidth]{
\begin{tikzpicture}[>=latex,line join=bevel,scale=0.6]
\node (0) at (39.5bp,22.0bp) [draw,circle, double,initial,initial text=] {$0$};
  \draw [->] (0) ..controls (37.396bp,53.795bp) and (38.044bp,62.0bp)  .. (39.5bp,62.0bp) .. controls (40.387bp,62.0bp) and (40.975bp,58.953bp)  .. (0);
  \definecolor{strokecol}{rgb}{0.0,0.0,0.0};
  \pgfsetstrokecolor{strokecol}
  \draw (9.0bp,54.5bp) node {$(t_{11} , t_{11})$};
  \draw [->] (0) ..controls (34.763bp,61.724bp) and (35.891bp,80.0bp)  .. (39.5bp,80.0bp) .. controls (42.404bp,80.0bp) and (43.702bp,68.166bp)  .. (0);
  \draw (9.0bp,72.5bp) node {$(t_{12} , t_{12})$};
  \draw [->] (0) ..controls (32.087bp,68.903bp) and (33.532bp,98.0bp)  .. (39.5bp,98.0bp) .. controls (44.675bp,98.0bp) and (46.449bp,76.119bp)  .. (0);
  \draw (9.0bp,90.5bp) node {$(t_{21} , t_{21})$};
  \draw [->] (0) ..controls (29.417bp,75.815bp) and (31.092bp,116.0bp)  .. (39.5bp,116.0bp) .. controls (47.054bp,116.0bp) and (49.174bp,83.563bp)  .. (0);
  \draw (9.0bp,108.5bp) node {$(t_{22} , t_{22})$};
\end{tikzpicture}
}
\subcaptionbox{Desired language $\desired$\label{fig:sim_desired}}[0.55\linewidth]{
\begin{tikzpicture}[>=latex,line join=bevel,scale=0.42]
\begin{scope}
  \node (1) at (175.0bp,117.0bp) [draw,circle,double,solid] {$1$};
  \node (0) at (22.0bp,86.0bp) [draw,circle,double,initial,initial text=] {$0$};
  \node (3) at (175.0bp,55.0bp) [draw,circle,double,solid] {$3$};
  \node (2) at (328.0bp,146.0bp) [draw,circle,double,solid] {$2$};
  \node (5) at (481.0bp,118.0bp) [draw,circle,double,solid] {$5$};
  \node (4) at (328.0bp,84.0bp) [draw,circle,double,solid] {$4$};
  \node (7) at (481.0bp,56.0bp) [draw,circle,double,solid] {$7$};
  \node (6) at (328.0bp,22.0bp) [draw,circle,double,solid] {$6$};
  \node (8) at (634.0bp,87.0bp) [draw,circle,double,solid] {$8$};
\end{scope}
\begin{scope}
  \draw [->] (0) ..controls (48.209bp,77.754bp) and (53.78bp,76.221bp)  .. (59.0bp,75.0bp) .. controls (87.127bp,68.419bp) and (119.56bp,63.008bp)  .. (3);
  \definecolor{strokecol}{rgb}{0.0,0.0,0.0};
  \pgfsetstrokecolor{strokecol}
  \draw (98.5bp,82.5bp) node {$(t_{21} , t_{21})$};
  \draw [->] (3) ..controls (200.07bp,42.18bp) and (206.16bp,39.703bp)  .. (212.0bp,38.0bp) .. controls (239.59bp,29.952bp) and (272.07bp,25.913bp)  .. (6);
  \draw (251.5bp,45.5bp) node {$(t_{22} , t_{22})$};
  \draw [->] (4) ..controls (373.45bp,91.063bp) and (411.63bp,97.717bp)  .. (444.0bp,106.0bp) .. controls (446.04bp,106.52bp) and (448.14bp,107.1bp)  .. (5);
  \draw (404.5bp,113.5bp) node {$(t_{12} , t_{12})$};
  \draw [->] (5) ..controls (528.53bp,108.37bp) and (572.27bp,99.508bp)  .. (8);
  \draw (557.5bp,116.5bp) node {$(t_{22} , t_{22})$};
  \draw [->] (4) ..controls (353.43bp,72.538bp) and (359.36bp,70.431bp)  .. (365.0bp,69.0bp) .. controls (392.77bp,61.958bp) and (425.24bp,58.729bp)  .. (7);
  \draw (404.5bp,76.5bp) node {$(t_{22} , t_{22})$};
  \draw [->] (6) ..controls (373.94bp,24.515bp) and (412.29bp,28.368bp)  .. (444.0bp,38.0bp) .. controls (446.76bp,38.837bp) and (449.56bp,39.857bp)  .. (7);
  \draw (404.5bp,45.5bp) node {$(t_{11} , t_{11})$};
  \draw [->] (7) ..controls (526.41bp,62.378bp) and (564.57bp,68.411bp)  .. (597.0bp,76.0bp) .. controls (599.04bp,76.477bp) and (601.13bp,77.002bp)  .. (8);
  \draw (557.5bp,83.5bp) node {$(t_{12} , t_{12})$};
  \draw [->] (1) ..controls (201.18bp,107.98bp) and (206.76bp,106.32bp)  .. (212.0bp,105.0bp) .. controls (240.06bp,97.945bp) and (272.49bp,92.278bp)  .. (4);
  \draw (251.5bp,112.5bp) node {$(t_{21} , t_{21})$};
  \draw [->] (3) ..controls (220.71bp,57.319bp) and (258.98bp,60.735bp)  .. (291.0bp,69.0bp) .. controls (293.47bp,69.637bp) and (295.99bp,70.4bp)  .. (4);
  \draw (251.5bp,76.5bp) node {$(t_{11} , t_{11})$};
  \draw [->] (0) ..controls (69.534bp,95.631bp) and (113.27bp,104.49bp)  .. (1);
  \draw (98.5bp,115.5bp) node {$(t_{11} , t_{11})$};
  \draw [->] (2) ..controls (375.94bp,137.23bp) and (419.53bp,129.25bp)  .. (5);
  \draw (404.5bp,145.5bp) node {$(t_{21} , t_{21})$};
  \draw [->] (1) ..controls (222.94bp,126.09bp) and (266.53bp,134.35bp)  .. (2);
  \draw (251.5bp,145.5bp) node {$(t_{12} , t_{12})$};
\end{scope}
\end{tikzpicture}
}
\subcaptionbox{Sensor Attack Model\label{fig:sim_ao}}[.15\linewidth]{
\begin{tikzpicture}[>=latex,line join=bevel,scale=0.6]
\node (0) at (42.0bp,22.0bp) [draw,circle, double,initial,initial text=] {$0$};
  \draw [->] (0) ..controls (39.896bp,53.795bp) and (40.544bp,62.0bp)  .. (42.0bp,62.0bp) .. controls (42.887bp,62.0bp) and (43.475bp,58.953bp)  .. (0);
  \definecolor{strokecol}{rgb}{0.0,0.0,0.0};
  \pgfsetstrokecolor{strokecol}
  \draw (12.0bp,59.5bp) node {$(t_{11} , \epsilon)$};
  \draw [->] (0) ..controls (37.263bp,61.724bp) and (38.391bp,80.0bp)  .. (42.0bp,80.0bp) .. controls (44.904bp,80.0bp) and (46.202bp,68.166bp)  .. (0);
  \draw (12.0bp,77.5bp) node {$(t_{12} , \epsilon)$};
  \draw [->] (0) ..controls (34.587bp,68.903bp) and (36.032bp,98.0bp)  .. (42.0bp,98.0bp) .. controls (47.175bp,98.0bp) and (48.949bp,76.119bp)  .. (0);
  \draw (12.0bp,95.5bp) node {$(t_{21} , t_{21})$};
  \draw [->] (0) ..controls (31.917bp,75.815bp) and (33.592bp,116.0bp)  .. (42.0bp,116.0bp) .. controls (49.554bp,116.0bp) and (51.674bp,83.563bp)  .. (0);
  \draw (12.0bp,113.5bp) node {$(t_{22} , t_{22})$};
\end{tikzpicture}
}

\par\bigskip
\subcaptionbox{Actuator Attack Model\label{fig:sim_ai}}[.35\linewidth]{
\begin{tikzpicture}[>=latex,line join=bevel,scale=0.55]
\node (1) at (188.5bp,91.499bp) [draw,ellipse,solid] {$1$};
  \node (0) at (39.5bp,59.499bp) [draw,circle, double,initial,initial text=] {$0$};
  \node (2) at (188.5bp,26.499bp) [draw,ellipse,solid] {$2$};
  \draw [->] (1) ..controls (167.45bp,79.965bp) and (161.35bp,77.272bp)  .. (155.5bp,75.499bp) .. controls (128.0bp,67.165bp) and (95.502bp,63.156bp)  .. (0);
  \definecolor{strokecol}{rgb}{0.0,0.0,0.0};
  \pgfsetstrokecolor{strokecol}
  \draw (116.0bp,82.999bp) node {$(t_{21} , t_{11}$};
  \draw [->] (0) ..controls (59.361bp,83.934bp) and (67.463bp,90.902bp)  .. (76.5bp,94.499bp) .. controls (103.77bp,105.35bp) and (137.8bp,102.37bp)  .. (1);
  \draw (116.0bp,109.0bp) node {$(t_{11} , t_{21})$};
  \draw [->] (0) ..controls (64.579bp,46.698bp) and (70.668bp,44.216bp)  .. (76.5bp,42.499bp) .. controls (104.32bp,34.311bp) and (137.29bp,30.246bp)  .. (2);
  \draw (116.0bp,49.999bp) node {$(t_{12} , t_{22})$};
  \draw [->] (2) ..controls (167.91bp,13.827bp) and (161.64bp,10.987bp)  .. (155.5bp,9.499bp) .. controls (121.38bp,1.2264bp) and (107.85bp,-6.304bp)  .. (76.5bp,9.499bp) .. controls (67.267bp,14.152bp) and (59.765bp,22.319bp)  .. (0);
  \draw (116.0bp,16.999bp) node {$(t_{22} , t_{12})$};
  \draw [->] (0) ..controls (35.659bp,91.062bp) and (36.828bp,99.499bp)  .. (39.5bp,99.499bp) .. controls (41.17bp,99.499bp) and (42.253bp,96.203bp)  .. (0);
  \draw (-3.5bp,91.0bp) node {$(t_{11} , t_{11})$};
  \draw [->] (0) ..controls (30.882bp,98.805bp) and (32.895bp,117.5bp)  .. (39.5bp,117.5bp) .. controls (44.867bp,117.5bp) and (47.202bp,105.16bp)  .. (0);
  \draw (-3.5bp,109.0bp) node {$(t_{12} , t_{12})$};
  \draw [->] (0) ..controls (26.033bp,105.8bp) and (28.59bp,135.5bp)  .. (39.5bp,135.5bp) .. controls (49.004bp,135.5bp) and (52.169bp,112.96bp)  .. (0);
  \draw (-3.5bp,127.0bp) node {$(t_{21} , t_{21})$};
  \draw [->] (0) ..controls (21.168bp,112.13bp) and (24.062bp,153.5bp)  .. (39.5bp,153.5bp) .. controls (53.37bp,153.5bp) and (57.115bp,120.11bp)  .. (0);
  \draw (-5.5bp,145.0bp) node {$(t_{22} , t_{22})$};
\end{tikzpicture}
}
\subcaptionbox{Supervisor \label{fig:sim_supervisor}}[.55\linewidth]{
\begin{tikzpicture}[>=latex,line join=bevel,scale=0.42]
\node (10) at (655.5bp,98.0bp) [draw,circle, double,solid] {$10$};
  \node (1) at (180.0bp,98.0bp) [draw,ellipse,solid] {$1$};
  \node (0) at (22.0bp,98.0bp) [draw,circle, double,initial,initial text=] {$0$};
  \node (3) at (180.0bp,40.0bp) [draw,circle, double,solid] {$3$};
  \node (2) at (180.0bp,156.0bp) [draw,circle, double,solid] {$2$};
  \node (5) at (338.0bp,174.0bp) [draw,circle, double,solid] {$5$};
  \node (4) at (338.0bp,98.0bp) [draw,circle, double,solid] {$4$};
  \node (7) at (496.0bp,98.0bp) [draw,ellipse,solid] {$7$};
  \node (6) at (338.0bp,22.0bp) [draw,circle, double,solid] {$6$};
  \node (9) at (496.0bp,40.0bp) [draw,circle, double,solid] {$9$};
  \node (8) at (496.0bp,156.0bp) [draw,circle, double,solid] {$8$};
  \draw [->] (0) ..controls (46.924bp,84.44bp) and (53.13bp,81.434bp)  .. (59.0bp,79.0bp) .. controls (88.792bp,66.646bp) and (124.0bp,55.689bp)  .. (3);
  \definecolor{strokecol}{rgb}{0.0,0.0,0.0};
  \pgfsetstrokecolor{strokecol}
  \draw (101.0bp,86.5bp) node {$(t_{21} , t_{21})$};
  \draw [->] (3) ..controls (204.94bp,26.676bp) and (211.07bp,24.324bp)  .. (217.0bp,23.0bp) .. controls (246.47bp,16.426bp) and (280.99bp,16.919bp)  .. (6);
  \draw (259.0bp,30.5bp) node {$(t_{22} , t_{22})$};
  \draw [->] (0) ..controls (46.924bp,111.56bp) and (53.13bp,114.57bp)  .. (59.0bp,117.0bp) .. controls (88.792bp,129.35bp) and (124.0bp,140.31bp)  .. (2);
  \draw (101.0bp,152.5bp) node {$(\epsilon , t_{11})$};
  \draw [->] (5) ..controls (386.9bp,168.43bp) and (432.75bp,163.21bp)  .. (8);
  \draw (417.0bp,176.5bp) node {$(t_{21} , t_{21})$};
  \draw [->] (4) ..controls (363.62bp,109.5bp) and (369.5bp,111.93bp)  .. (375.0bp,114.0bp) .. controls (411.71bp,127.8bp) and (422.14bp,127.62bp)  .. (459.0bp,141.0bp) .. controls (461.25bp,141.82bp) and (463.57bp,142.69bp)  .. (8);
  \draw (417.0bp,148.5bp) node {$(\epsilon , t_{12})$};
  \draw [->] (4) ..controls (388.46bp,98.0bp) and (436.92bp,98.0bp)  .. (7);
  \draw (417.0bp,110.5bp) node {$(t_{22} , t_{12})$};
  \draw [->] (6) ..controls (384.68bp,17.219bp) and (425.17bp,15.453bp)  .. (459.0bp,23.0bp) .. controls (461.69bp,23.6bp) and (464.42bp,24.411bp)  .. (9);
  \draw (417.0bp,30.5bp) node {$(\epsilon , t_{11})$};
  \draw [->] (7) ..controls (540.43bp,98.0bp) and (588.92bp,98.0bp)  .. (10);
  \draw (575.0bp,110.5bp) node {$(\epsilon , t_{22})$};
  \draw [->] (1) ..controls (224.27bp,98.0bp) and (272.98bp,98.0bp)  .. (4);
  \draw (259.0bp,110.5bp) node {$(\epsilon , t_{21})$};
  \draw [->] (8) ..controls (541.76bp,143.22bp) and (582.84bp,130.85bp)  .. (617.0bp,117.0bp) .. controls (619.72bp,115.9bp) and (622.51bp,114.68bp)  .. (10);
  \draw (575.0bp,152.5bp) node {$(t_{22} , t_{22})$};
  \draw [->] (9) ..controls (541.76bp,52.775bp) and (582.84bp,65.153bp)  .. (617.0bp,79.0bp) .. controls (619.72bp,80.103bp) and (622.51bp,81.315bp)  .. (10);
  \draw (575.0bp,86.5bp) node {$(\epsilon , t_{12})$};
  \draw [->] (0) ..controls (72.457bp,98.0bp) and (120.92bp,98.0bp)  .. (1);
  \draw (101.0bp,110.5bp) node {$(t_{21} , t_{11})$};
  \draw [->] (2) ..controls (228.9bp,161.57bp) and (274.75bp,166.79bp)  .. (5);
  \draw (259.0bp,176.5bp) node {$(\epsilon , t_{12})$};
  \draw [->] (3) ..controls (225.83bp,52.586bp) and (266.95bp,64.881bp)  .. (301.0bp,79.0bp) .. controls (303.84bp,80.179bp) and (306.77bp,81.492bp)  .. (4);
  \draw (259.0bp,86.5bp) node {$(\epsilon , t_{11})$};
  \draw [->] (4) ..controls (362.92bp,84.44bp) and (369.13bp,81.434bp)  .. (375.0bp,79.0bp) .. controls (404.79bp,66.646bp) and (440.0bp,55.689bp)  .. (9);
  \draw (417.0bp,86.5bp) node {$(t_{22} , t_{22})$};
  \draw [->] (2) ..controls (205.94bp,145.17bp) and (211.66bp,142.94bp)  .. (217.0bp,141.0bp) .. controls (253.86bp,127.62bp) and (264.29bp,127.8bp)  .. (301.0bp,114.0bp) .. controls (303.41bp,113.09bp) and (305.89bp,112.12bp)  .. (4);
  \draw (259.0bp,148.5bp) node {$(t_{21} , t_{21})$};
\end{tikzpicture}
}
\caption{Example supervisors resilient to sensor and actuator attacks\label{fig:sim_example}}
\end{figure*}







\section{Conclusions}
\label{sec:conclusion}

We studied the supervisory control of discrete-event systems under sensor and actuator attacks. Both the system and the attacks are mathematically modeled by finite-state transducers (FST). We discussed the advantage of using FSTs to model attacks in capturing history dependency, composing multiple attacks, and imposing constraints on attacks. Then we proposed theorems and algorithms to design resilient supervisors for given attacks in three cases: only sensor attacks, only actuator attacks, and both sensor and actuator attacks. Finally, we implemented the algorithms with computer programs and demonstrated their applicability in case studies. Our work extended previous studies on attack-resilient supervisory control and provided a new framework based on FSTs.

\bibliographystyle{IEEEtran}
\bibliography{Zotero,SecureControl,CPSL@DukePapers,ref,yu}

\begin{IEEEbiography}
[{\includegraphics[width=1in]{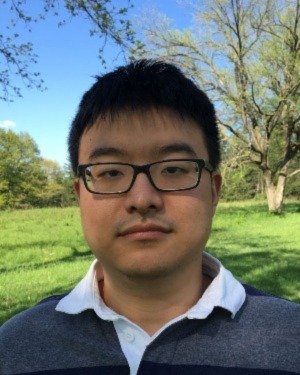}}]
{Yu Wang}{\space}
is an Assistant Professor at the Department of Mechanical and Aerospace Engineering at the University of Florida (UF) and the Group Lead of Autonomous and Connected Vehicles of the UF Transportation Institute. He was a Postdoctoral Researcher at the Department of Electrical and Computer Engineering at Duke University. He received his Ph.D. in Mechanical Engineering and M.S. in Statistics and Mathematics from the University of Illinois at Urbana-Champaign (UIUC). His research focuses on assured autonomy, cyber-physical systems, machine learning, and formal methods. This work on statistical verification of hyperproperties for cyber-physical systems was selected as one of the Best Paper Finalists of the ACM SIGBED International Conference on Embedded Software (EMSOFT) in 2019.
\end{IEEEbiography}

\begin{IEEEbiography}[{\includegraphics[width=1in,height=1.25in,clip,keepaspectratio]{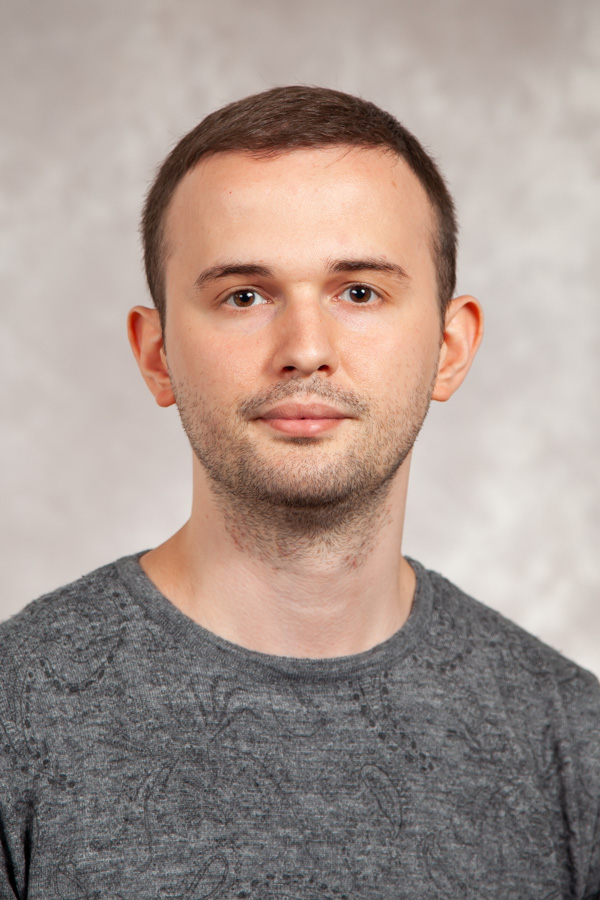}}]{Alper Kamil Bozkurt}{\space} received the B.S. and M.S. degrees in computer engineering from Bogazici University, Turkey, in 2015 and 2018, respectively. He is currently a Ph.D. candidate in the Department of Computer Science at Duke University. His research interests lie at the intersection of machine learning, control theory, and formal methods. In particular, he focuses on developing learning-based algorithms that synthesize provably safe and reliable controllers for cyber-physical systems.
\end{IEEEbiography}

\begin{IEEEbiography}[{\includegraphics[width=1in,height=1.25in,clip,keepaspectratio]{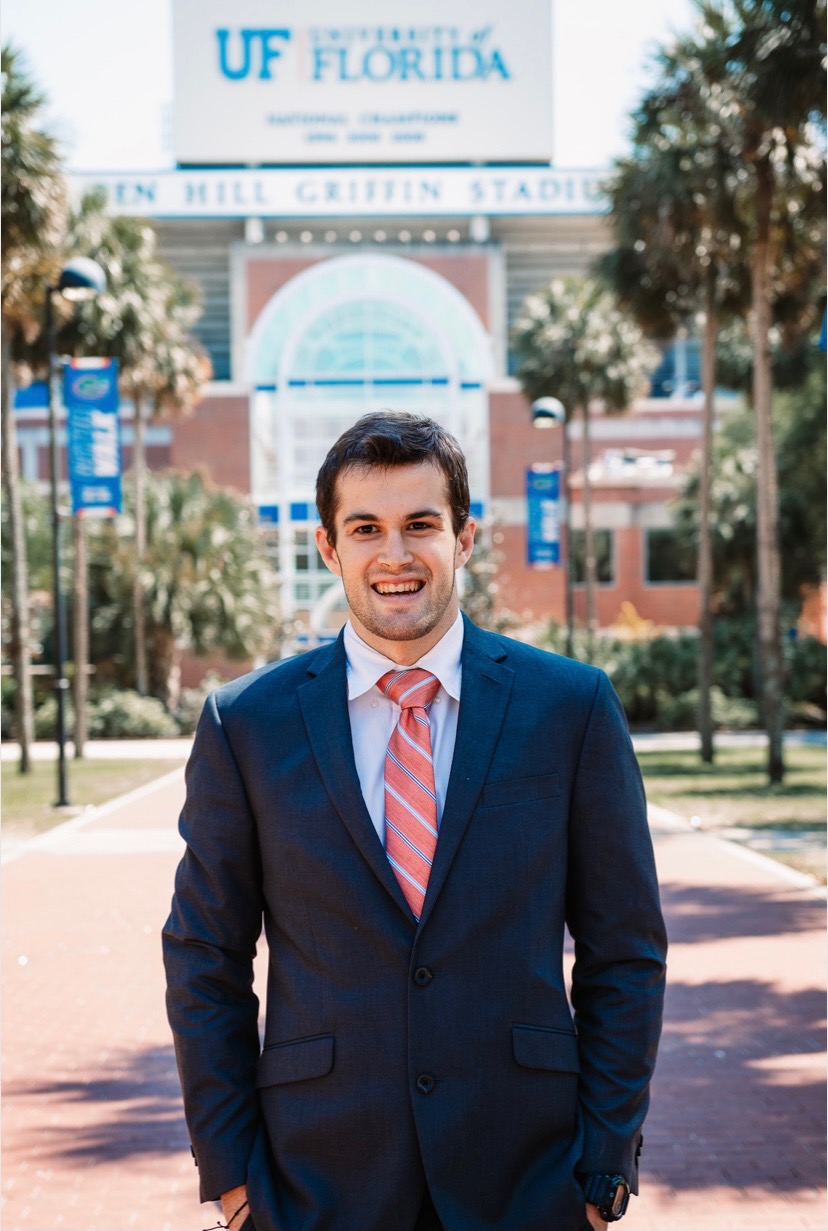}}]{Nathan Smith}{\space} received his Bachelor of Science in Aerospace Engineering and Mathematics from the University of Florida in 2021. He is currently pursuing a Ph.D. in Aerospace Engineering under Dr. Yu Wang. He is from Jupiter, Florida. His research interests include applications of Game Theory and  Reinforcement Learning to model attacks on Cyber-Physical systems.
\end{IEEEbiography}

\begin{IEEEbiography}[{\includegraphics[trim={60pt 230pt 75pt 40pt},width=1in,height=1.25in,clip,keepaspectratio]{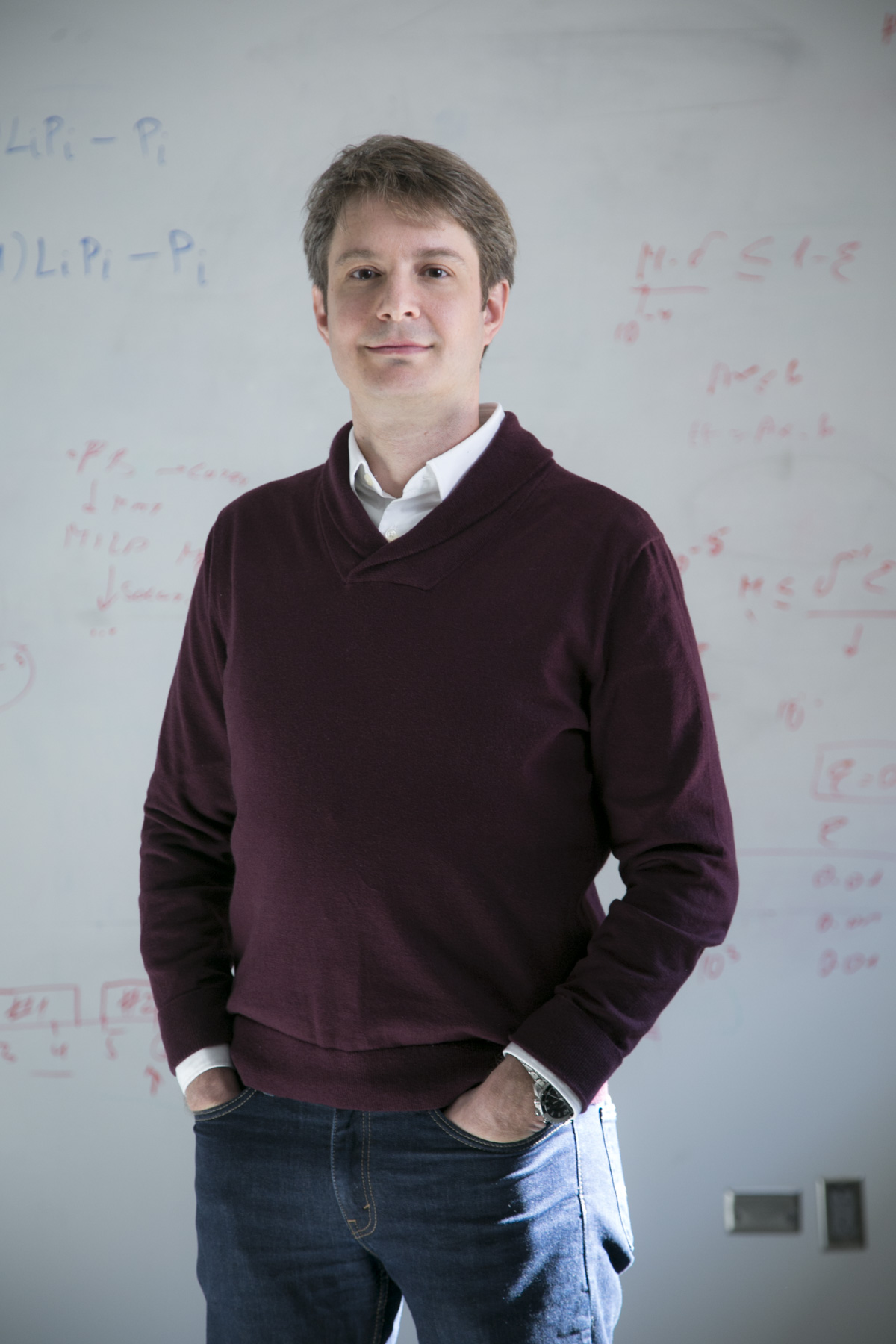}}]{Miroslav Pajic}{\space} received the Dipl. Ing. and M.S. degrees in electrical engineering from the University of Belgrade, Serbia, in 2003 and 2007, respectively, and the M.S. and Ph.D. degrees in electrical engineering from the University of Pennsylvania, Philadelphia, in 2010 and 2012, respectively.

He is currently the Dickinson Family Associate Professor in Department of Electrical and Computer Engineering at Duke University. He also holds secondary appointments in the Computer Science, and Mechanical Engineering and Material Science Departments. His research interests focus on the design and analysis of high-assurance cyber-physical systems with varying levels of autonomy and human interaction, at the intersection of (more traditional) areas of learning and controls, AI, embedded systems, formal methods, and robotics. 
  
Dr. Pajic received various awards including the ACM SIGBED Early-Career Award, IEEE TCCPS Early-Career Award, NSF CAREER Award, ONR Young Investigator Program Award, ACM SIGBED Frank Anger Memorial Award, Joseph and Rosaline Wolf Best Dissertation Award from Penn Engineering, IBM Faculty Award, as well as seven Best Paper and Runner-up Awards at top cyber-physical systems venues. 
\end{IEEEbiography}

\end{document}